\documentclass[aps,prl,twocolumn,superscriptaddress,floatfix,letter]{revtex4-1}
\usepackage[bookmarks=true,colorlinks,linkcolor=OrangeRed,urlcolor=NavyBlue,citecolor=RoyalBlue]{hyperref}
\usepackage{amsmath,amssymb,color,comment,ulem,physics}
\usepackage[makeroom]{cancel}
\usepackage[caption=false]{subfig}
\usepackage{mathrsfs}
\usepackage{graphicx}
\usepackage{subfig}
\usepackage[countmax]{subfloat}
\usepackage[english]{babel}
\usepackage[dvipsnames]{xcolor}

\usepackage{braket}

\newcommand{\rme}{{\rm e}}

\definecolor{mypurple}{rgb}{0.49,0.18,0.56}
\definecolor{mygold}{rgb}{0.93,0.49,0.13}
\definecolor{mygreen}{rgb}{0,0.5,0}
\definecolor{myblue}{rgb}{0,0,0.75}
\definecolor{mymagenta}{cmyk}{0,1,0,0.12}
\definecolor{mygray}{rgb}{0.5,0.5,0.5}

\usepackage{umoline}

\definecolor{mypink1}{rgb}{0.858, 0.188, 0.478}

\begin{document}

\title{Gauge-Symmetry Violation Quantum Phase Transition in Lattice Gauge Theories}
\author{Maarten Van Damme}
\affiliation{Department of Physics and Astronomy, University of Ghent, Krijgslaan 281, 9000 Gent, Belgium}

\author{Jad C.~Halimeh}
\affiliation{INO-CNR BEC Center and Department of Physics, University of Trento, Via Sommarive 14, I-38123 Trento, Italy}

\author{Philipp Hauke}
\affiliation{INO-CNR BEC Center and Department of Physics, University of Trento, Via Sommarive 14, I-38123 Trento, Italy}

\begin{abstract}
Gauge symmetry plays a key role in our description of subatomic matter. The vanishing photon mass, the long-ranged Coulomb law, and asymptotic freedom are all due to gauge invariance. 
Recent years have seen tantalizing progress in the microscopic reconstruction of gauge theories in engineered quantum simulators. Yet, many of these are plagued by a fundamental question: When gauge symmetry is only approximate in the quantum device, do we actually quantum-simulate a gauge theory? 
Here, we answer this question in the affirmative for a paradigm gauge theory akin to quantum electrodynamics. 
Analytically, we derive a renormalized gauge symmetry that is at least exponentially accurate. Further, numerically computing the phase diagram in the thermodynamic limit, we find that the long-distance behavior of the gauge theory is only compromised upon reaching a sharp quantum phase transition. 
This behavior is enabled by an energy penalty term, which lends a mass to the Higgs boson to which the coherent gauge breaking couples. Our results not only lend validity to ongoing gauge-theory quantum simulations, they also probe the fundamental question of how gauge symmetry could emerge in nature.
\end{abstract}
\date{\today}
\maketitle

\begin{figure}[ht!]
	\centering
	\includegraphics[width=.45\textwidth]{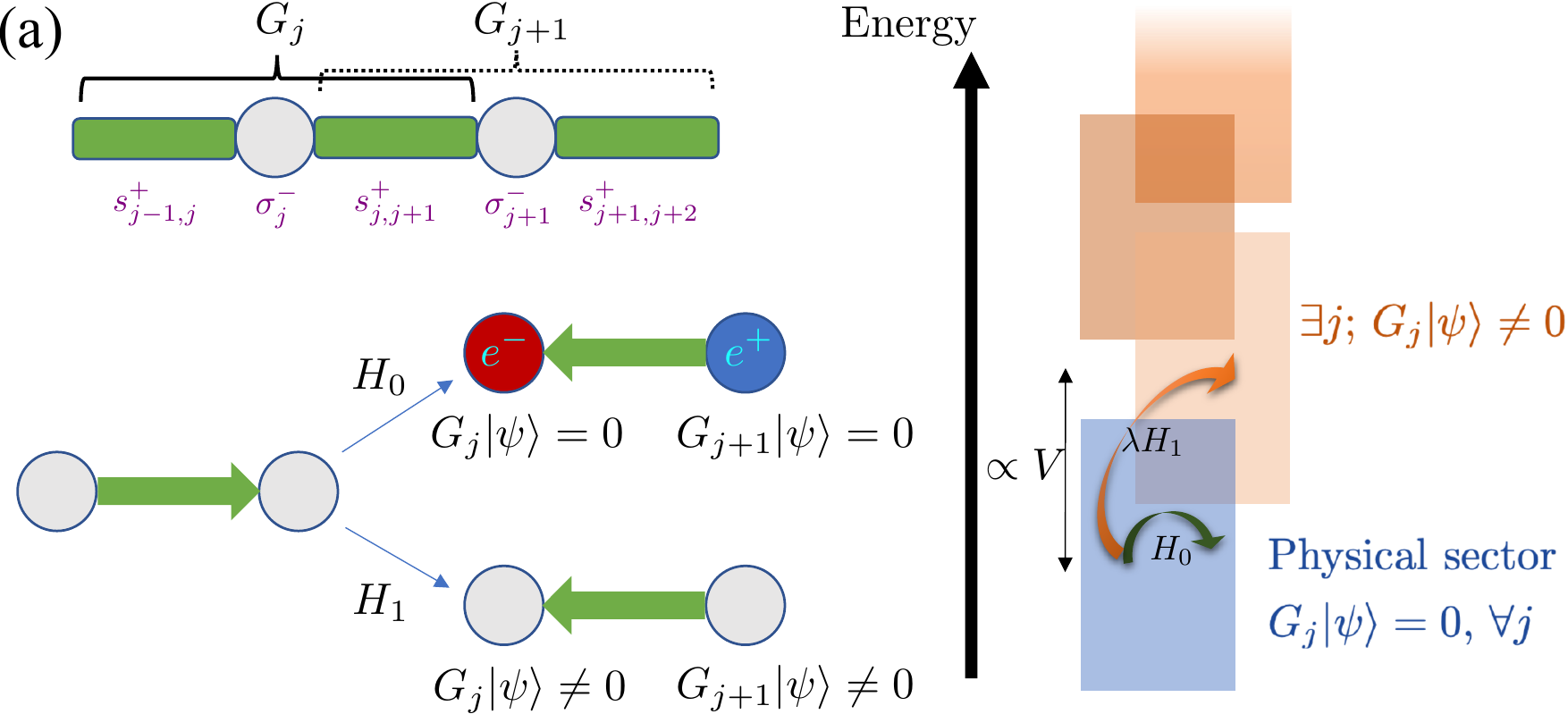}\\
	\vspace{0.15cm}
	\includegraphics[width=.23\textwidth]{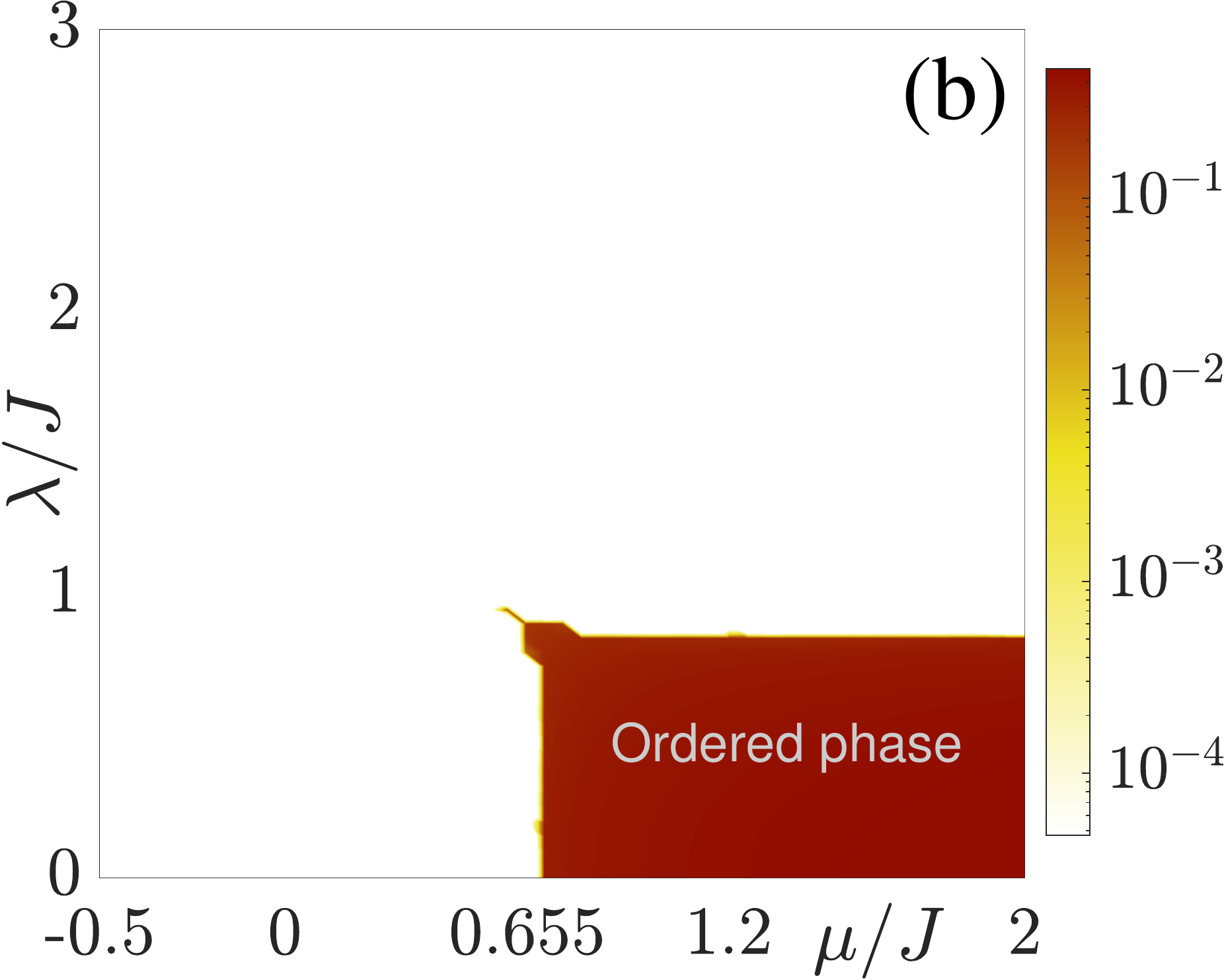}\quad
	\includegraphics[width=.23\textwidth]{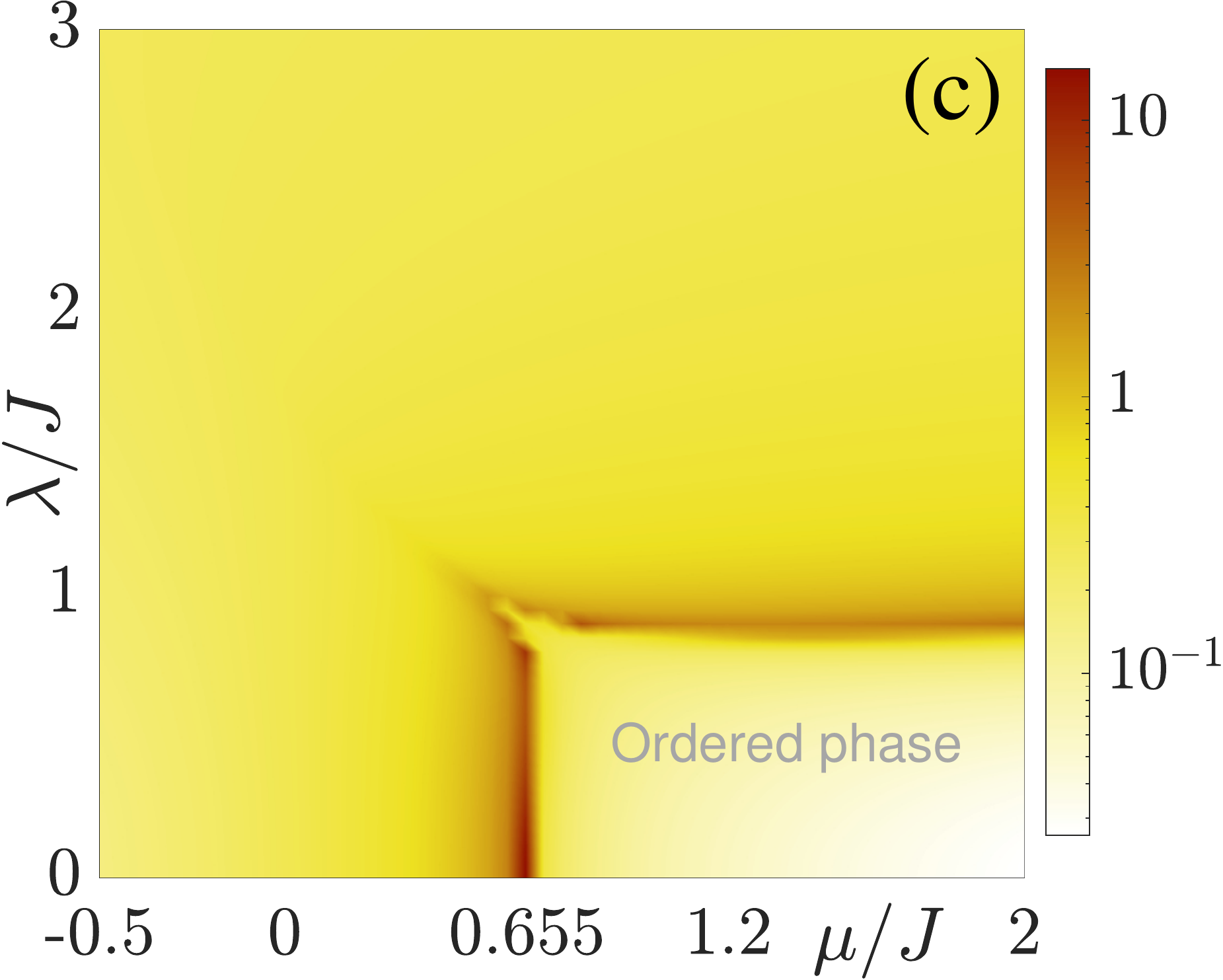}\\
	\includegraphics[width=.23\textwidth]{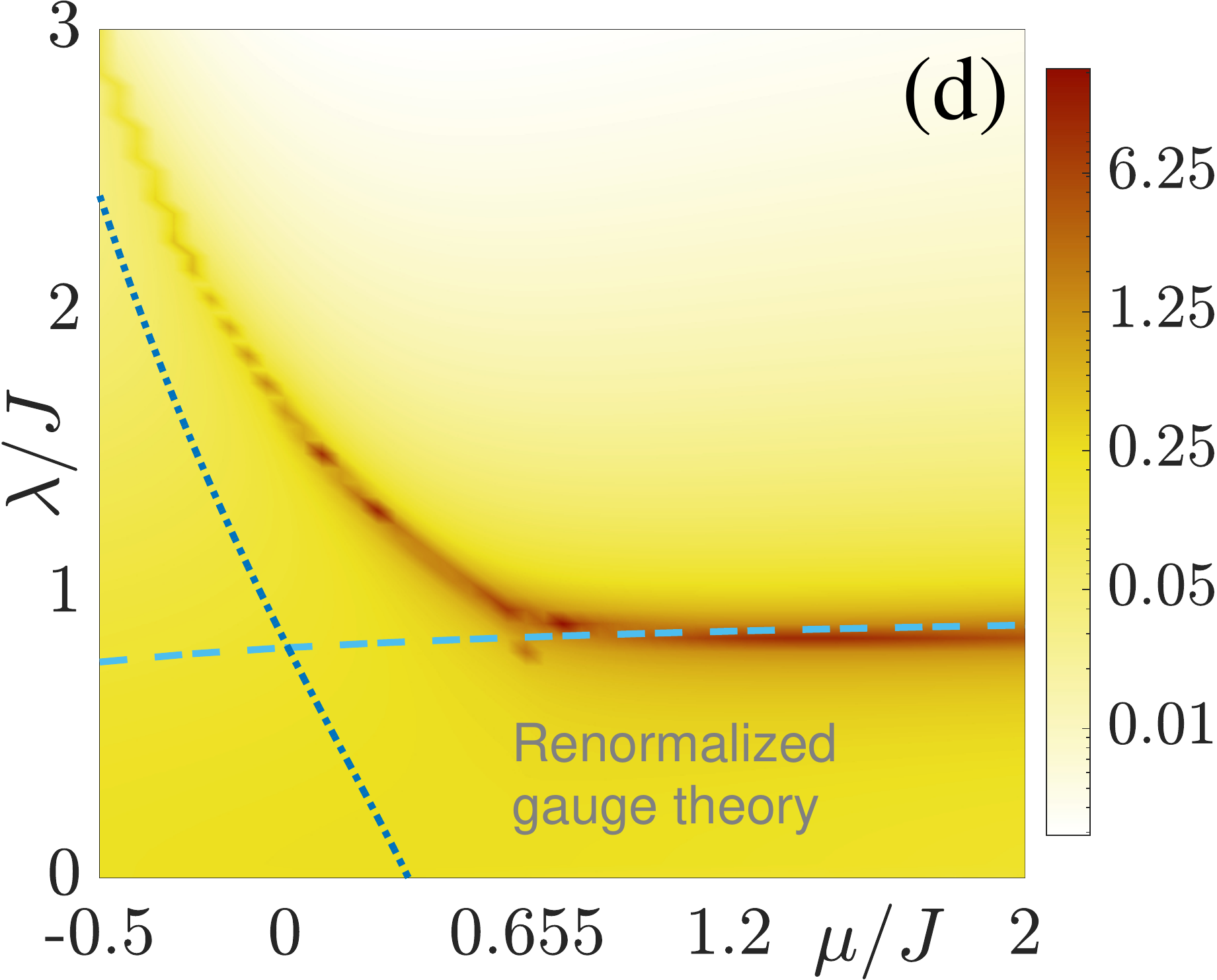}\quad
	\includegraphics[width=.23\textwidth]{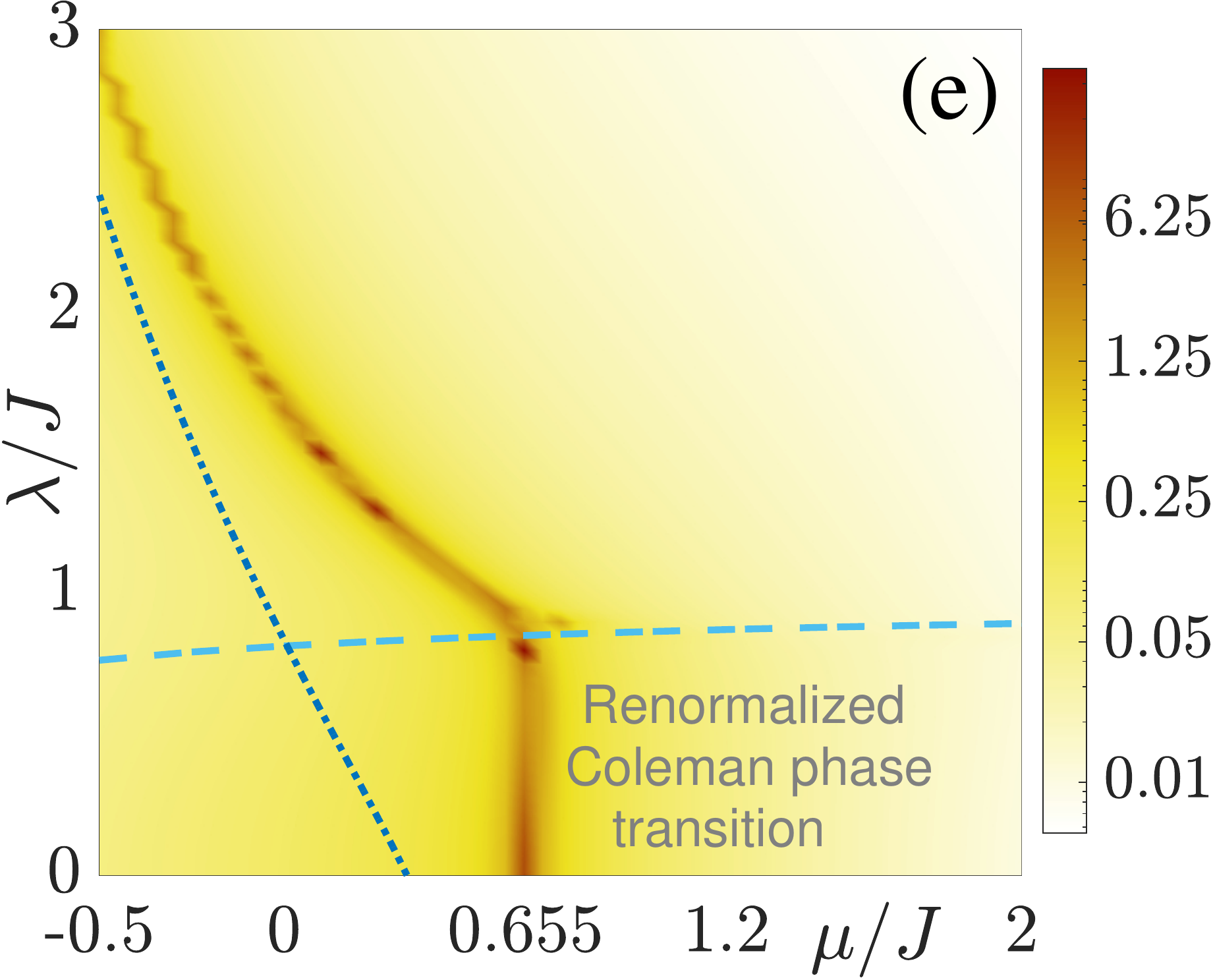}
	\caption{(a) 
	An Abelian lattice gauge theory, ideally described by Hamiltonian $H_0$ and gauge-symmetry generators $G_j$ that are conserved quantities. 
	The target theory lives in the `physical sector' $G_j\ket{\psi}_\mathrm{phys}=0$, $\forall j$. 
	In realistic quantum simulators, gauge symmetry-breaking terms $\lambda H_1$ drive the dynamics outside this sector. 
	A protection term $VH_G=V\sum_jG_j^2$ renders gauge-breaking processes energetically unfavorable. 
	(b-e) A rich quantum phase diagram emerges with a renormalized gauge theory at low $\lambda/V$, as revealed by (b) the order parameter of Coleman's phase transition, (c) its connected correlator, and (d,e) the fidelity susceptibility. 
	Predictions from a Gutzwiller mean-field decoupling between gauge and matter fields (dashed lines in d,e) show good qualitative agreement. 
}
	\label{fig:PD}
\end{figure}

The role of gauge symmetry in our description of subatomic physics cannot be overstated, as it is the cornerstone of the Standard Model of Particle Physics, which unites such important theories as quantum electrodynamics (QED) and quantum chromodynamics (QCD) 
\cite{Weinberg_book,Gattringer_book}. 
Gauge symmetry describes the invariance of physical observables against local transformations, and in fundamental descriptions of nature it is postulated as a natural law. 
As a consequence, in subatomic physics traditionally only a minority of works is concerned with clarifying the physical consequences of breakings of gauge symmetry, e.g., \cite{Foerster1980,Poppitz2008,Wetterich2017,Bass2020}, though valuable insights can be drawn from the context of topological phases of matter 
\cite{Hastings2005,Sachdev2018}. 
Currently, this issue is gaining a reinvigorated importance as we are witnessing an explosion of implementations in engineered quantum devices, such as those based on trapped ions \cite{Martinez2016,Kokail2019}, ultracold atoms \cite{Bernien2017,Goerg2019,Schweizer2019,Mil2019,Yang2020}, and superconducting qubits \cite{Klco2018}. 
The aim of these experiments is to solve the complex dynamics of gauge theories \cite{Wiese_review,Zohar2015,Dalmonte2016,MariCarmen2019}, but in many of them gauge symmetry is only approximately fulfilled. 

Of particular importance in this context is QED in 1+1D known as the Schwinger model \cite{Coleman1976} or derivates of it such as quantum-link model (QLM) variations \cite{Chandrasekharan1997,Wiese_review}, where an Abelian $\mathrm{U}(1)$ gauge symmetry is embodied in the well-known Gauss's law. 
These are among the simplest models hosting a gauge field as well as dynamical charges in the form of electronic and positronic matter, and they share many features with the much more complex gauge theory of QCD in 3+1D, such as the Schwinger mechanism of particle--antiparticle creation and a topological $\theta$-angle. 
Moreover, living in 1+1D, these theories lend themselves for benchmarking against efficient classical methods such as those based on tensor networks \cite{Dalmonte2016,MariCarmen2019}. 
These beneficial properties have made this class of models the workhorse and standard benchmark for gauge-theory quantum simulation \cite{Martinez2016,Bernien2017,Klco2018,Kokail2019,Mil2019,Yang2020}. 

Various theory works have shown that such quantum simulators can acquire small violations of gauge symmetry \cite{Zohar2011,Zohar2012,Zohar2013,Banerjee2013,Hauke2013,Stannigel2014,Kuehn2014,Yang2016,Dutta2017,Kuno2017,Negretti2017,Barros2019,Halimeh2020a,Halimeh2020d,Halimeh2020e,Lamm2020},
and a recent experiment has successfully certified a small degree of gauge violation \cite{Yang2020}. 
Despite all this progress, a lingering issue has been that even slight breakings of gauge symmetry can have dramatic consequences such as unconventional prethermalization out of equilibrium \cite{Halimeh2020b,Halimeh2020c} or a photon with nonzero mass \cite{Tu2004} (though a vanishing photon mass may still appear in a renormalized theory \cite{Foerster1980,Hastings2005,Poppitz2008}). 
Thus, the compelling question remains whether error-prone quantum simulators can realize the physics of a gauge theory, particularly when the aim is to study infrared properties at long distances.

Here, we tackle this issue by computing the phase diagram of a QLM version of the Schwinger model in the presence of coherent gauge-breaking errors.  
To this end, we exploit a state-of-the-art iMPS tensor-network code \cite{MPSKit}, allowing us to directly treat the thermodynamic limit.  
As we demonstrate as one of our main results, adding an energy-penalty term opens up an extended phase where the long-wavelength physics of the desired gauge theory is retained. This phase is governed by an emergent gauge symmetry that we show to be at least exponentially accurate (in gauge-breaking over gauge-protecting strength), and whose low-order perturbative form we explicitly derive. 
Only at sizeable gauge-breaking strength does a paramagnetic high-error phase arise, which is separated by a clear phase transition.
Further, we explain our findings through a fictitious Higgs boson with heavy mass and a Gutzwiller mean-field analysis. 
This latter analysis also establishes interesting connections between the considered theory and coupled Ising--XY models, which are known for their rich phase diagrams and unconvential universality \cite{Hasenbusch2005,Granato1991}. 
As these results demonstrate, not only can quantum simulators benefit from a broad parameter range within which they can reproduce the entire physics of the ideal model, the gauge breaking itself also gives rise to an enriched phase diagram that is worthwhile exploring in its own right.

\textbf{\emph{Model.---}} In this work, we consider an Abelian $\mathrm{U}(1)$ lattice gauge theory in the QLM formalism \cite{Wiese_review,Chandrasekharan1997}, sketched in Fig.~\ref{fig:PD}a. 
After a Jordan-Wigner transformation of fermionic matter (electrons and positrons) to hard-core bosons or qubits as well as a staggered particle-hole transformation, the model reads \cite{Hauke2013,Yang2016}
\begin{align}\label{eq:Ham0}
H_0=-J\sum_{j=1}^L\big(\sigma^-_j s^+_{j,j+1}\sigma^-_{j+1}+\text{H.c.}\big)+\frac{\mu}{2}\sum_j\sigma^z_j.
\end{align}
Here, $L$ is the system size, where our infinite matrix product state (iMPS) calculations \cite{MPSKit} work effectively in the thermodynamic limit $L\to\infty$ (see Appendix). Further, $\sigma^{\{x,y,z\}}_j$ are Pauli matrices on site $j$, with $\sigma^+_j$ ($\sigma^-_j$) representing the creation (annihilation) operator of the hard-core bosonic matter field and $\sigma^{z}_j$ counting the occupation number. Accordingly, $\mu$ represents the rest mass of electrons and positrons. 
Their kinetic energy is coupled via the term $\propto J$ to a QLM gauge field. This field can be seen as a coarse electric field with only the two eigenvalues $\pm 1/2$, and is represented by the spin-$1/2$ matrices $s^{\{x,y,z\}}_{j,j+1}$ living on links $(j,j+1)$ between matter sites. 
The term $\propto J$ thus describes the generation of an electron-positron pair with the according adaption of the electric field (see Fig.~\ref{fig:PD}a). 

Gauge symmetry in this model is embodied in the conservation of the Gauss's-law operators 
\begin{align}
G_j=\frac{(-1)^j}{2}\big[\sigma^z_j+2\big(s^z_{j,j+1}+s^z_{j-1,j}\big)+1\big]\,. 
\end{align}
Ideally, the generators are conserved quantities, $[H_0,G_j]=[G_j,G_l]=0$, $\forall j,l$, which split the gauge theory into separate sectors. Without loss of generality, we focus on the so-called `physical sector' with the Gauss's law $G_j\ket{\psi}_\mathrm{phys}=0$, $\forall j$. 

In a realistic quantum device, errors may arise that explicitly break gauge symmetry \footnote{Note that according to Elitzur's theorem, a gauge symmetry cannot be spontaneously broken.}. We model these as systematic coherent errors described by 
\begin{align}
\label{eq:H1}
\lambda H_1= \lambda\sum_j\big[c_1\big(\sigma^-_j \sigma^-_{j+1}+\text{H.c.}\big)+2c_2s^x_{j,j+1}\big],
\end{align}
describing unassisted matter tunneling and gauge-field flipping; see Fig.~\ref{fig:PD}a for an illustration of the latter (if not stated otherwise, $c_1=c_2=1$). 
Here, we are particularly interested in answering whether the ideal physics can be restored with the help of an energy-protection term
\begin{align}
V H_G=V\sum_j G_j^2\,.
\end{align}
In the past, it has been shown numerically for small systems, analytically, and in an experiment that such terms can reduce the degree of gauge violation 
\cite{Zohar2011,Zohar2012,Zohar2013,Banerjee2013,Hauke2013,Stannigel2014,Kuehn2014,Yang2016,Dutta2017,Kuno2017,Negretti2017,Barros2019,Halimeh2020a,Halimeh2020d,Halimeh2020e,Lamm2020}. 
Here, we demonstrate that in the thermodynamic limit it leads to distinct `gauge-protected' phases, similar to what has been observed in pure gauge theories in 2+1D in the context of topologically ordered phases of matter \cite{Hastings2005}.

\textbf{\emph{Phase diagram.---}} 
In the physical sector, the ideal theory $H_0$ hosts an analog of Coleman's quantum phase transition in the Schwinger model at topological $\theta$-angle $\pi/2$: at the critical point $\mu/J\approx 0.655$, the system goes from a charge-proliferated phase to one with a broken global $\mathrm{Z}_2$ symmetry where the electric field passes unhindered through the system \cite{Coleman1976,Pichler2016}. The order parameter heralding this transition is the staggered magnetization $\mathcal{M}_z=\lvert\sum_j(-1)^js^z_j\rvert/L$, which maps to a homogeneous electric field.  

A major open question is whether this phase transition persists in the presence of gauge-breaking errors, as these may dramatically modify the symmetries of the Hamiltonian. 
Here, we settle this question using iMPS, a powerful numerical technique that works directly in the thermodynamic limit, to compute the ground state of the full Hamiltonian 
\begin{align}
H=H_0+\lambda H_1+VH_G. 
\end{align}

As Fig.~\ref{fig:PD}b illustrates for a gauge-protection strength $V=2$, the order parameter persists at a large value over a broad region of $\lambda/V$ (see also Fig.~\ref{fig:Peierls}). Coleman's phase transition extends almost vertically towards increasing $\lambda$, and the order parameter only drops abruptly to zero once $\lambda/J=\mathcal{O}(1)$. 
This behavior is reflected in a strong feature of the connected correlator $\sum_{j,\ell}(-1)^{j+\ell}\big(\langle s^z_js^z_\ell\rangle-\langle s^z_j\rangle\langle s^z_\ell\rangle\big)/L$, plotted in Fig.~\ref{fig:PD}c, indicating a divergent correlation length and thus signaling a quantum phase transition. 

To exclude the possibility of a sharp crossover, we further consider the fidelity susceptibilities defined as \cite{Gu2010}
\begin{align}
\chi_\eta = \frac{\partial \bra{\psi_{\mu,\lambda,V}} }{\partial \eta} \big(\mathbb{I} - \ket{\psi_{\mu,\lambda,V}}\bra{\psi_{\mu,\lambda,V}}\big) \frac{\partial \ket{\psi_{\mu,\lambda,V}}}{\partial \eta}\,,
\end{align}
where $H(\mu,\lambda,V) \ket{\psi_{\mu,\lambda,V} } = E(\mu,\lambda,V) \ket{\psi_{\mu,\lambda,V}}$ and where $\eta=\mu,J$. 
Divergences in $\chi_\eta$ showcase quantum phase transitions without the need to know the order parameter (for the transition to be detected, the parameter direction $\eta$ along which the derivative is taken must not lie parallel to the transition line). 

Indeed, as seen in Fig.~\ref{fig:PD}d,e (see Appendix for cuts through the phase diagram), we find a strong divergence of $\chi_\eta$ exactly at the line where the order parameter drops. 
Furthermore, an additional divergence emerges within the parameter region where the vertical transition line (emanating from the ideal Coleman's phase transition) and the horizontal transition line (where the order parameter drops) meet. As these findings show, there is a low-gauge-error phase with the same physics and ordering behavior as the ideal theory, which is separated from a distinct phase at large gauge error. 

\begin{figure}[ht!]
	\centering
	\includegraphics[width=.23\textwidth]{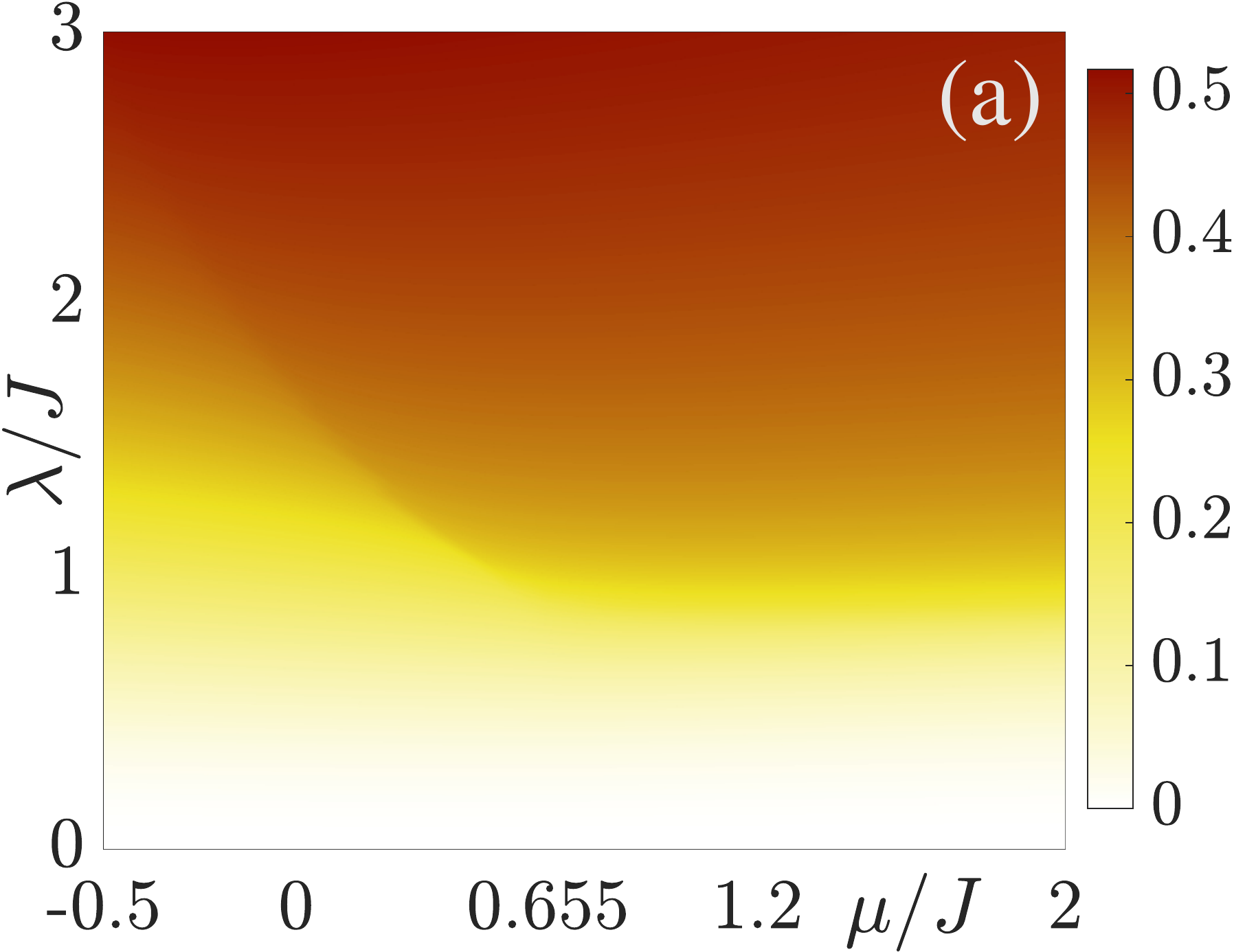}\quad
	\includegraphics[width=.23\textwidth]{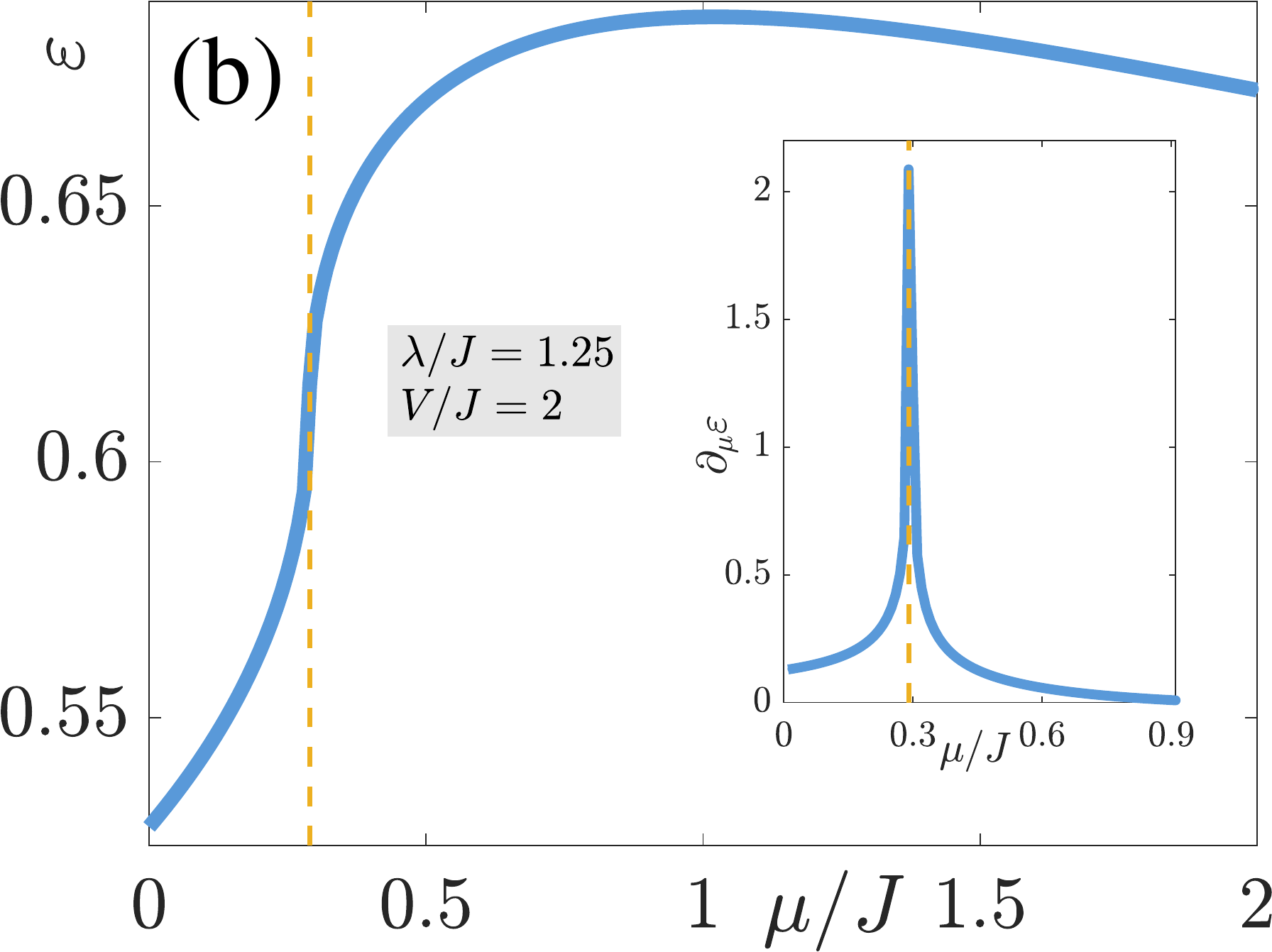}\\
	\includegraphics[width=.23\textwidth]{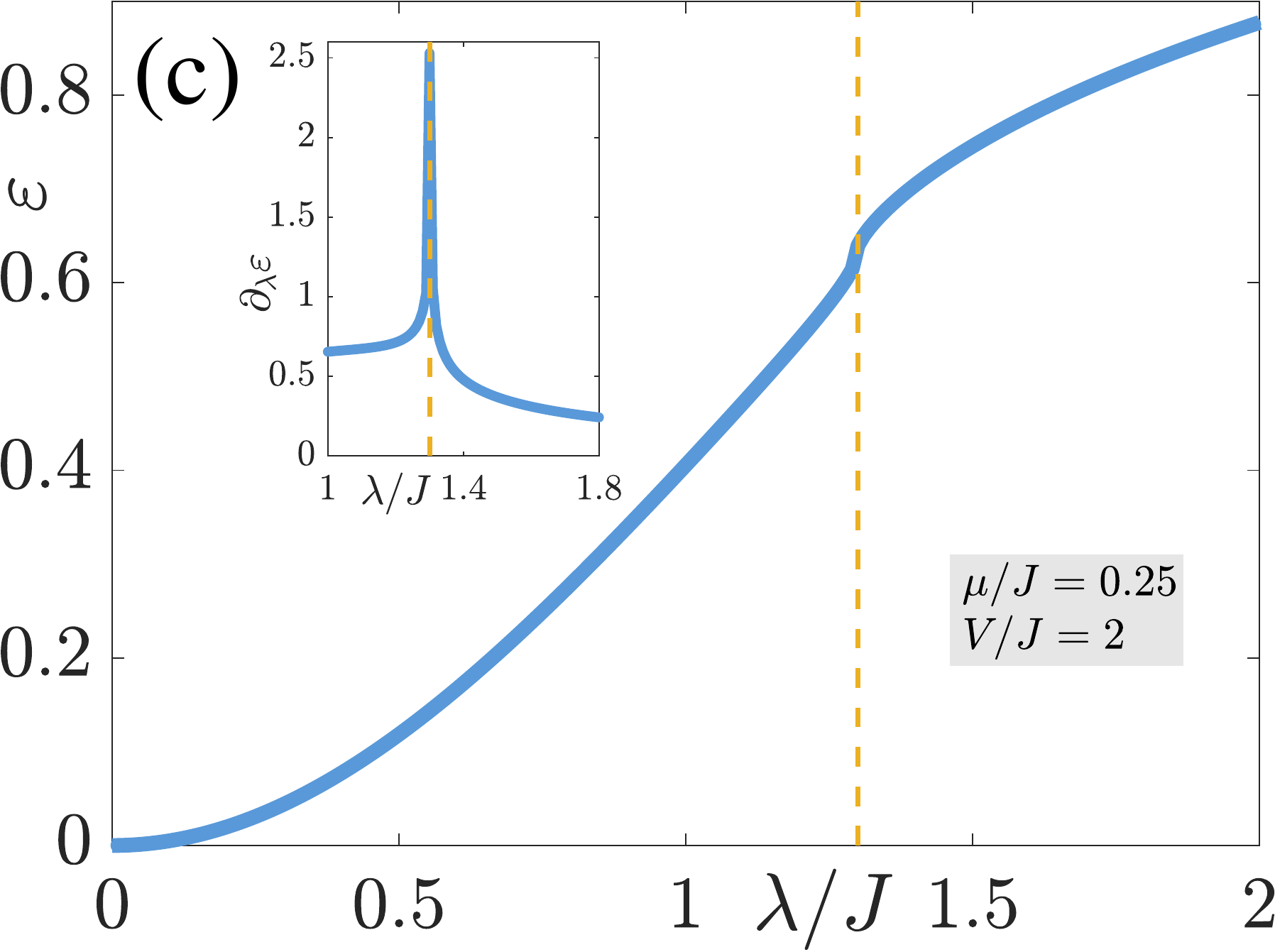}\quad
	\includegraphics[width=.23\textwidth]{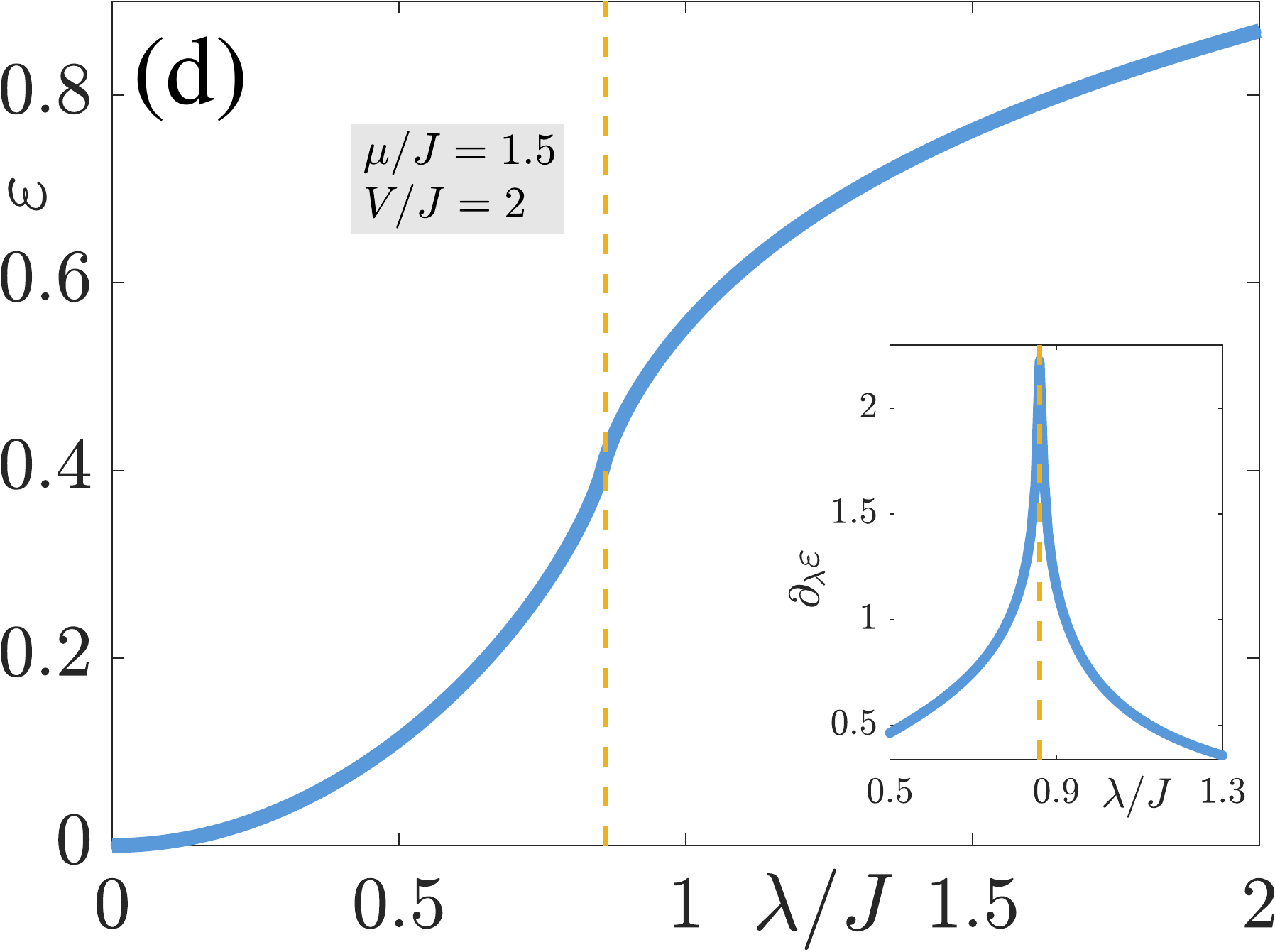}
	\caption{Renormalized gauge symmetry. 
		(a) Bare gauge violation $\varepsilon=\braket{H_G}/L$ across parameter space. 
		(b-d) Cuts along fixed $\lambda$ and $\mu$. 
		For low $\lambda/V$, $\braket{H_G}/L$ increases smoothly, while it experiences a nonanalyticity at the quantum phase transition. 
	}
	\label{fig:Gausslaw}
\end{figure}

\textbf{\emph{Gauge-symmetry-violating phase transition.---}} 
Not only is the low-gauge-error region continuously connected to the ideal theory $H_0$, it is characterized by an emergent gauge symmetry. Before deriving the corresponding generator, we plot in Fig.~\ref{fig:Gausslaw} the gauge violation $\varepsilon=\braket{H_G}/L$, which quantifies the deviation from the ideal `bare' gauge symmetry. 
As $H_G\ket{\psi}=0$ if and only if $G_j\ket{\psi}=0$, $\forall j$, the physical sector of the gauge theory coincides with the zero-eigenvalue sector of $H_G$. 
Even more, $\varepsilon$ can rigorously be connected to a mean-square displacement across gauge sectors as well as to the decrease in the overlap between two states that differ only by a gauge transformation \cite{Halimeh2020f,Halimeh2020g}.

A first indication of a drastic change in the gauge-symmetry behavior is visible in Fig.~\ref{fig:Gausslaw}: at points where the order parameter and fidelity susceptibility indicate a phase transition as a function of $\lambda$ (but not at the extension of Coleman's phase transition), $\varepsilon$ has clear nonanalyticities. Below these points, in contrast, it behaves in a smooth manner. This feature enables us to derive a renormalized symmetry generator $\tilde{H}_G=\rme^{-S} H_G \rme^{S}$, obtained by a unitary Schrieffer--Wolff transformation \cite{Bravyi2011} whose generator can be expanded in orders of $\lambda/V$, $S=\sum_k S_k$, see Appendix. 
Truncating the expansion at order $K$, the commutator between $\tilde{H}_G$ and $H/V$ is at most of $\mathcal{O}((\lambda/V)^{K+1})$, meaning that this transformation constructs an emergent symmetry of $H$ order by order (assuming convergence of the Schrieffer--Wolff transformation, see below). 
The same transformation defines renormalized Gauss's-law generators $\tilde{G}_j$ and a renormalized physical sector $\ket{\tilde{\psi}}_\mathrm{{phys}}$. 

\begin{figure}[ht!]
	\centering
	\includegraphics[width=.45\textwidth]{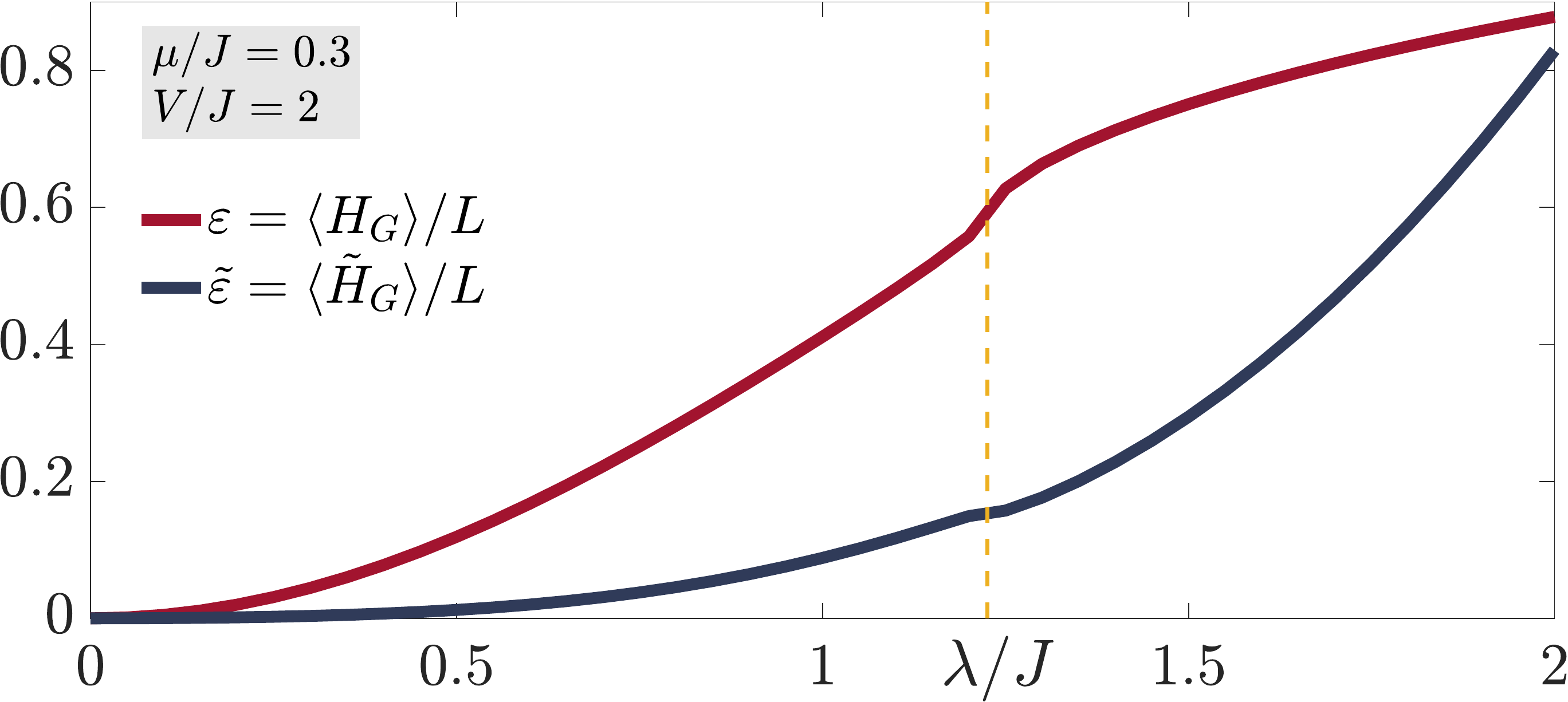}\\
	\caption{
		Dressed gauge symmetry. 
		The perturbative expansion of $\tilde{\varepsilon}=\braket{\tilde{H}_G}/L$ up to $\mathcal{O}((\lambda/V)^2)$ is systematically suppressed as compared to the bare gauge violation ${\varepsilon}=\braket{{H}_G}/L$. 
	}
	\label{fig:SW}
\end{figure}

Explicitly, up to $\mathcal{O}((\lambda/V)^2)$ we obtain 
\begin{align}
\braket{\tilde{H}_G}=\braket{H_G} + \frac{\lambda}{V}\braket{H_1} + \frac{1}{2}\frac{\lambda^2}{V^2}\braket{h_{1j}^2}\,,
\end{align}
where we defined $H_1=\sum_j h_{1j}$ as a sum of local contributions. 
As is illustrated in Fig.~\ref{fig:SW}, the dressed gauge violation $\tilde{\varepsilon}=\braket{\tilde{H}_G}/L$ is much diminished as compared to the bare gauge violation $\varepsilon=\braket{{H}_G}/L$. 

Importantly, since the perturbation strength is usually extensive, in a general many-body theory rigorous proofs of a convergence of this perturbative expansion require a dominating energy scale $V$ that scales as the system size \cite{Bravyi2011,Abanin2017}. 
We can, however, make rigorous statements using recent results that hold for operators with integer spectrum, as is the case for $H_G$. For sufficiently large $V$, the commutator between $\tilde{H}_G$ and $H$ can then be proven to be upper bounded by an exponentially small function \cite{Abanin2017,Halimeh2020e}, see Appendix. 
In principle, in the thermodynamic limit even an exponentially small breaking of a symmetry can still drive a system into a completely different phase. 
As is often the case for such rigorous mathematical proofs, however, our numerical iMPS calculations show that the exponentially small upper bound is still an overestimation. 
Only at the nonanalyticity encountered at $\lambda/V=\mathcal{O}(1)$ is a smooth deformation of the gauge symmetry no longer possible. Above it, there is a phase that has no remnance of the underlying gauge theory.

\textbf{\emph{Higgs field.---}} 
To shed more physical light on the energy penalty that $VH_G$ gives to gauge violations, we can reinterpret its role as follows.   
Consider for simplicity only the gauge-breaking term involving the gauge field, i.e., $c_1=0$ in Eq.~\eqref{eq:H1}. Such a term is often represented as coupled to an additional Higgs-like field $\phi$ \cite{Poppitz2008,Kuno2015,Bazavov2015,HeitgerPhDThesis}, 
\begin{equation}
	\lambda H_1^{\mathrm{Higgs}} = 2\lambda \sum_j  \phi_j^\dagger s_{j,j+1}^+ \phi_{j+1} + \mathrm{h.c.}\,.
\end{equation}
This field does not actually need to be implemented in the quantum simulator, but can be useful in discussing the physics of the gauge breaking. 
The full theory is then described by the Hamiltonian 
\begin{equation}
H ^{\mathrm{Higgs}}=H_0+\lambda H_1^{\mathrm{Higgs}} + V H_G
\end{equation}
and a modified Gauss’s law 
\begin{equation}
G_j^{\mathrm{Higgs}}=G_j + \phi_j^\dagger \phi_j - N_{\phi,j}^{(0)}\,,
\end{equation}
which commutes with $H ^{\mathrm{Higgs}}$. Here, $N_{\phi,j}^{(0)}$ is an integer constant chosen such that $G_j^{\mathrm{Higgs}}$ acting on the full wave function (comprising matter, gauge, and Higgs field) returns zero for all $j$. 
Inserting $G_j= G_j^{\mathrm{Higgs}}-(\phi_j^\dagger \phi_j - N_{\phi,j}^{(0)})$ into $H_G$, we obtain 
\begin{equation}
H ^{\mathrm{Higgs}}=H_0+\lambda H_1^{\mathrm{Higgs}}+2V\sum_j (\phi_j^\dagger \phi_j - N_{\phi,j}^{(0)})^2\,.
\end{equation}
At large $\lambda/V$, the kinetic energy for the Higgs field, described by $\lambda H_1^{\mathrm{Higgs}}$, leads to strong fluctuations of $\phi_j$. 
Instead, when $V$ dominates, it generates a potential that renders the Brout--Englert--Higgs mode massive \cite{Bazavov2015} and forces its occupation close to $N_{\phi,j}^{(0)}$. As a consequence, at small $\lambda/V$ the contribution of the Higgs field to $G_j^{\mathrm{Higgs}}$ is small, and $G_j$ is only slightly deformed, as we have seen in our analysis above.

\textbf{\emph{Gutzwiller analysis.---}}
As seen in Figs.~\ref{fig:PD} and~\ref{fig:Gausslaw}, the $\mathrm{U}(1)$ QLM under gauge-breaking and -protecting terms shows a rich phase diagram. 
Several of its features can be predicted by a Gutzwiller-type mean-field decoupling of gauge (g) and matter (m) fields, assuming the ground state to separate as $\ket{\psi_0}=\ket{\psi_0}_\mathrm{g}\otimes \ket{\psi_0}_\mathrm{m}$. 
That is clearly not a good Ansatz in the gauge-symmetric phase, as gauge- and matter-fields are intrinsically tied to each other. Nevertheless, it predicts qualitatively well the type of phase transitions that occur at increased $\lambda/V$.  

As detailed in the Appendix, this Ansatz yields the coupled Hamiltonians 
\begin{align}
	\label{eq:GWg}
		H^{\text{GW},g}=&\sum_j\big(2Vs^z_{j-1,j}s^z_{j,j+1}+\gamma_z s^z_{j,j+1}+\gamma_x s^x_{j,j+1}\big),
\end{align}
with $\gamma_x =[2\lambda-J(\langle\sigma^-_j\sigma^-_{j+1}\rangle+\mathrm{c.c.})$ and $\gamma_z =V(2+\langle\sigma^z_j\rangle + \langle\sigma^z_{j+1}\rangle)$, and 
\begin{align}
	\label{eq:GWm}
H^\mathrm{GW,m}&=\sum_j\big(-J_j^\mathrm{eff} \sigma^-_j \sigma^-_{j+1} + \text{H.c.}\big)+\delta_z \sum_j\sigma^z_j ,
\end{align}
with $J_j^\mathrm{eff}=-J \braket{s^+_{j,j+1}}+\lambda$ and 
$\delta_z=\frac{\mu}{2}+\frac{V}{2}\big[2\big(\braket{s^z_{j,j+1}}+\braket{s^z_{j-1,j}}\big)+1\big]$. 
Within this Ansatz, the system is thus described by an Ising antiferromagnet in transverse and longitudinal fields [Eq.~\eqref{eq:GWg}] coupled to an $XY$ model in a staggered transverse field [Eq.~\eqref{eq:GWg}, after a staggered rotation on the matter fields]. Such coupled Ising--XY models are famous for their rich phase diagrams. For example, they emerge as effective descriptions for fully-frustrated spin models, where they have enabled strong insights into the nature of phase transitions including unconventional critical behavior \cite{Hasenbusch2005,Granato1991}.  
Moreover, $\braket{s^+_{j,j+1}}$ can become staggered, $\mathcal{M}_{x}=(\braket{s^{x,z}_{2n-1,2n}}-\braket{s^{x,z}_{2n,2n+1}})/2\neq 0$ while $\braket{s^y_{j,j+1}}=0$, yielding $J_j^\mathrm{eff}=\bar{J}^\mathrm{eff}\pm \mathcal{M}_{x}$ alternatingly for $j$ even and odd, with $\bar{J}^\mathrm{eff}$ the average value (see Appendix). 
The result indicates a dimerization of the gauge-matter coupling term, i.e., a Peierls transition similar to what has been found in closely related models that do not have a gauge symmetry \cite{Cuadra2018}. 

Moreover, these Hamiltonians immediately hand us analytic predictions for phase transitions in limiting cases, in addition to Coleman's transition at $\mu/J\approx 0.655$ and $\lambda=0$. At $\mu=\infty$, the matter fields are frozen and we can set $\braket{\sigma_j^z}=-1$ and $\langle\sigma^-_j\sigma^-_{j+1}\rangle=0$. The gauge-field Hamiltonian in Eq.~\eqref{eq:GWg} then becomes the antiferromagnetic Ising model in a transverse field, which has a quantum phase transition at $\lambda=V/2$. Indeed, this is the transition point we find in this limit. 
This phase transition can persist to $\mu<\infty$ where $\gamma_z\neq 0$: 
contrary to ferromagnetic interactions, a homogeneous longitudinal field drives the antiferromagnetic Ising model to a paramagnetic phase only if it is sufficiently large. 
The antiferromagnetic XY model in a staggered transverse field $\delta_z$ but no homogeneous field has a phase transition 
at $\delta_z=0$ \cite{dutta2015quantum}. 
For $\lambda\to\infty$, we can set $\braket{s^z_{j,j+1}}=0$, and thus expect the transition to occur at ${\mu}=-V$, which again agrees with the numerical results. 

One can solve Eqs.~\eqref{eq:GWg} and~\eqref{eq:GWm} self-consistently to obtain the mean-field phase transitions away from these limiting points. The results are the dashed lines in Fig.~\ref{fig:PD}d,e, which display good qualitative agreement with the numerical prediction of the gauge symmetry-violating phase transition. The lines cross in a parameter region where also the extension of Coleman's phase transition lies. In that region, the system will become strongly correlated and the mean-field approach is no longer adequate, leading to a of shift the transition lines.

\textbf{\emph{Conclusion.---}} 
In conclusion, we find that an energy penalty term can protect a gauge theory up to considerable strengths of coherent gauge breaking. 
Since the analytic derivations of a renormalized gauge symmetry based on \cite{Abanin2017,Halimeh2020e} also hold for other Abelian symmetries, for global symmetries, and in higher dimensions, we expect the same to hold for a manifold of other scenarios. In this sense, our results expand on findings such as those of Ref.~\cite{Hastings2005} on pure gauge theories in 2+1D.  
Importantly, for a typical local error term, the generator of the perturbatively constructed approximate gauge symmetry is a local few-body observable, which can be accessed and certified in quantum simulation experiments. 
These results bode extremely well for ongoing experimental efforts to quantum simulate gauge theories. 

Even more, the interplay of ideal gauge theory, gauge symmetry breaking, and gauge protection generates an extremely rich phase diagram that is worthwhile to explore in its own right. Especially intriguing appear the connections to coupled XY--Ising models \cite{Hasenbusch2005,Granato1991} and systems with Peierls phase transitions \cite{Cuadra2018}.

\textbf{\emph{Acknowledgements.---}} 
We thank Marcello Dalmonte for helpful discussions and Zhang Jiang, Subir Sachdev, and Torsten Zache for useful comments. 
We acknowledge support by Provincia Autonoma di Trento, the ERC Starting Grant StrEnQTh (Project-ID 804305), Q@TN --- Quantum Science and Technology in Trento, ERC grants No 715861 (ERQUAF) and 647905 (QUTE)), and from Research Foundation Flanders (FWO) via grant GOE1520N.

\appendix

\section{iMPS method}

Our simulations were done within the formalism of matrix product states \cite{mps1,mps2,mps3}, a set of states that can efficiently approximate the ground state of local one-dimensional Hamiltonians \cite{faithfulmps}. One considerable advantage of this method is that we can efficiently work directly in the thermodynamic limit.

We find an approximate ground state by variationally minimizing the energy within our set of states. This procedure was done in practice using a mix of the VUMPS \cite{vumps} algorithm and gradient descent on Riemannian manifolds as described in Ref.~\cite{hauru2020riemannian}. While VUMPS performs rather well in the first few iterations, gradient descent has performance advantages closer to convergence.

\section{Additional numerical results}

\begin{figure}[ht!]
	\centering
	\includegraphics[width=.23\textwidth]{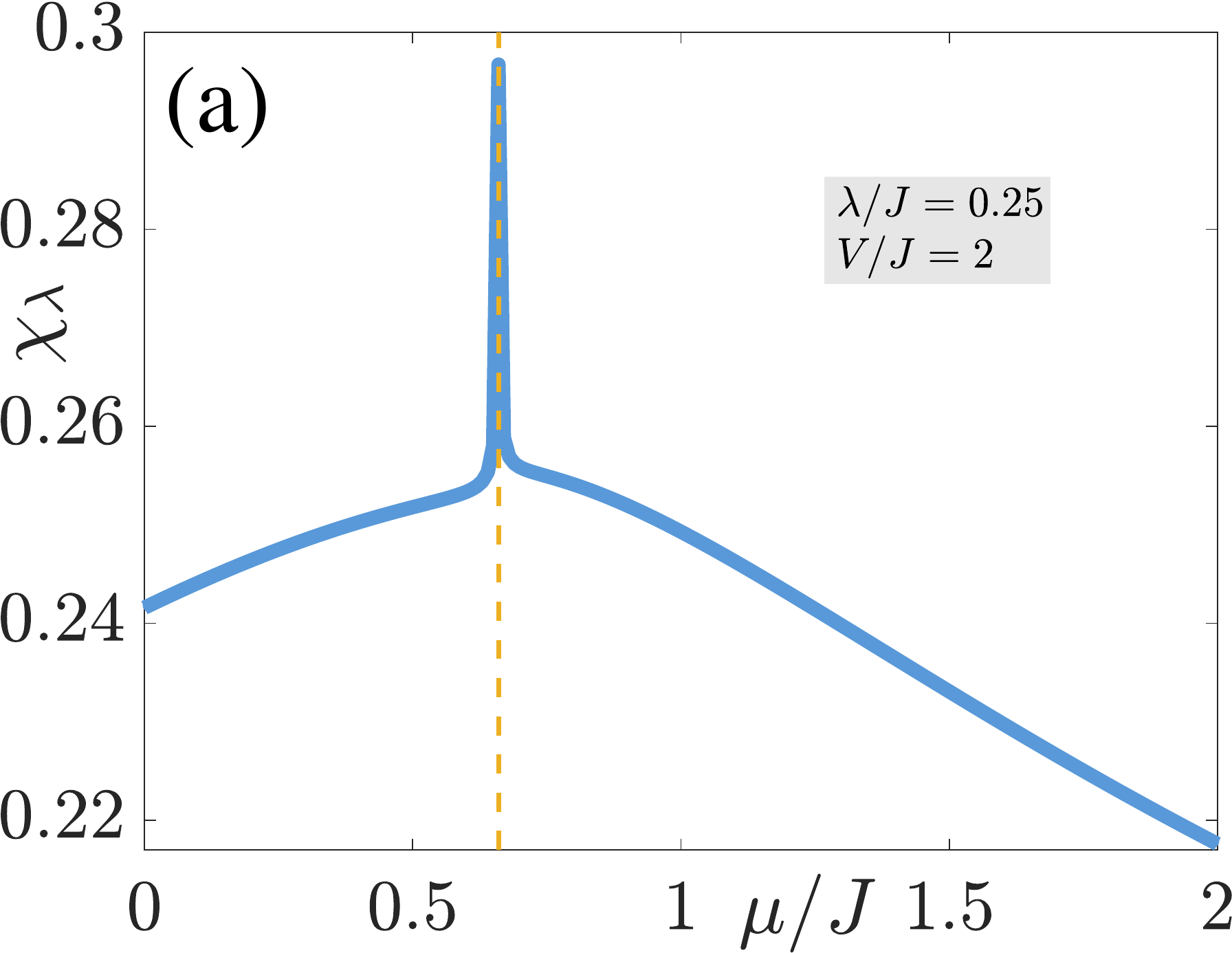}\quad
	\includegraphics[width=.23\textwidth]{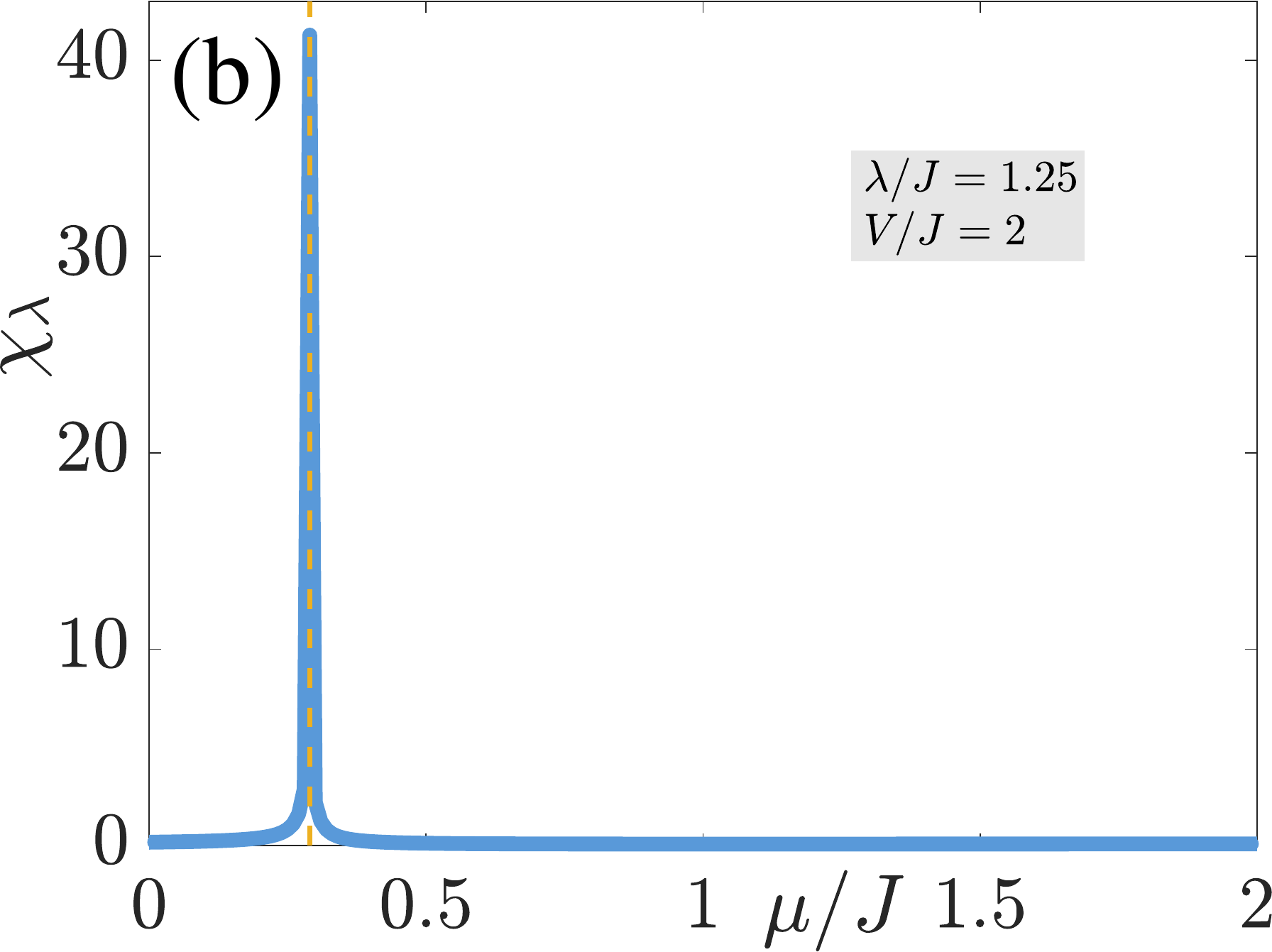}\\
	\includegraphics[width=.23\textwidth]{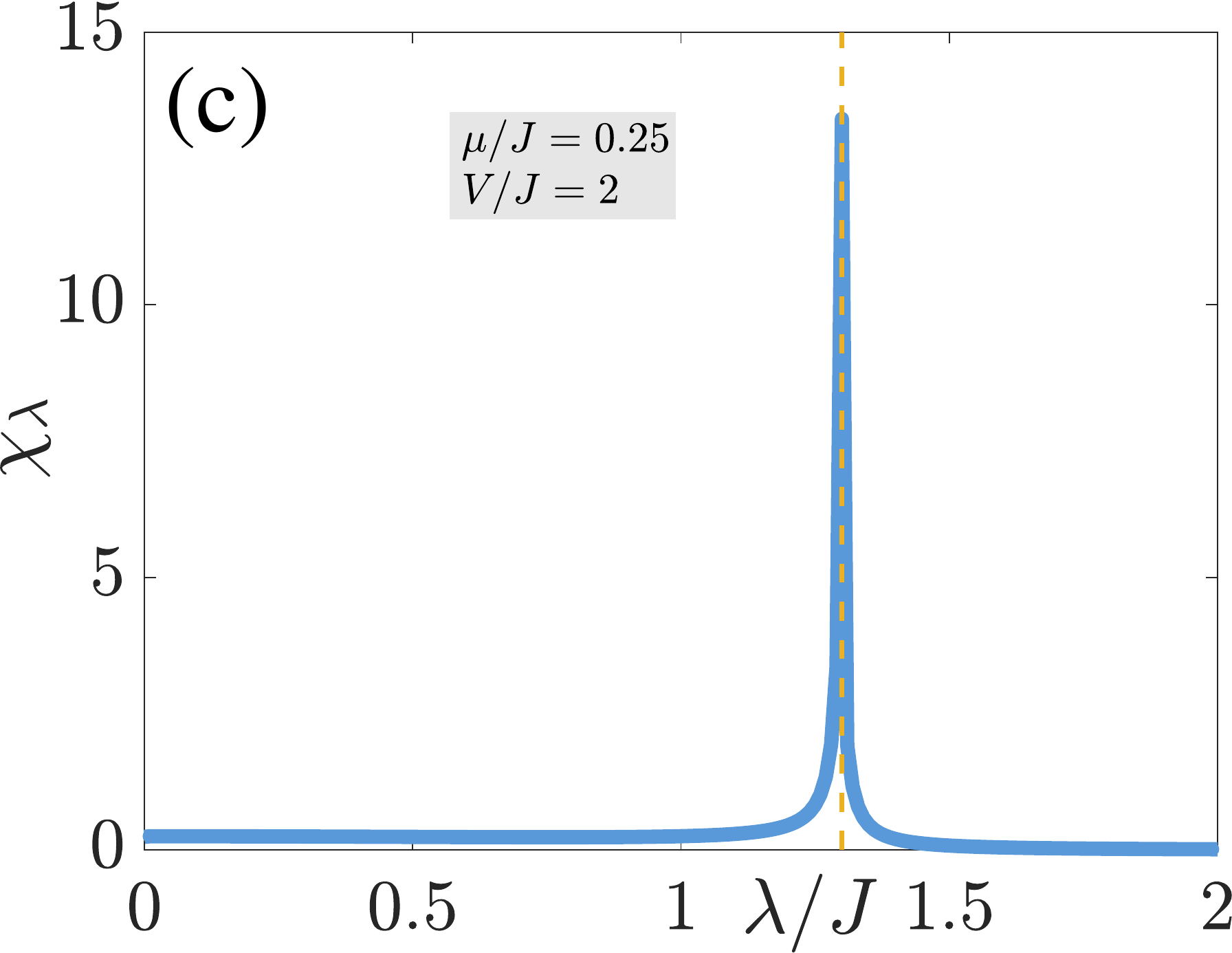}\quad
	\includegraphics[width=.23\textwidth]{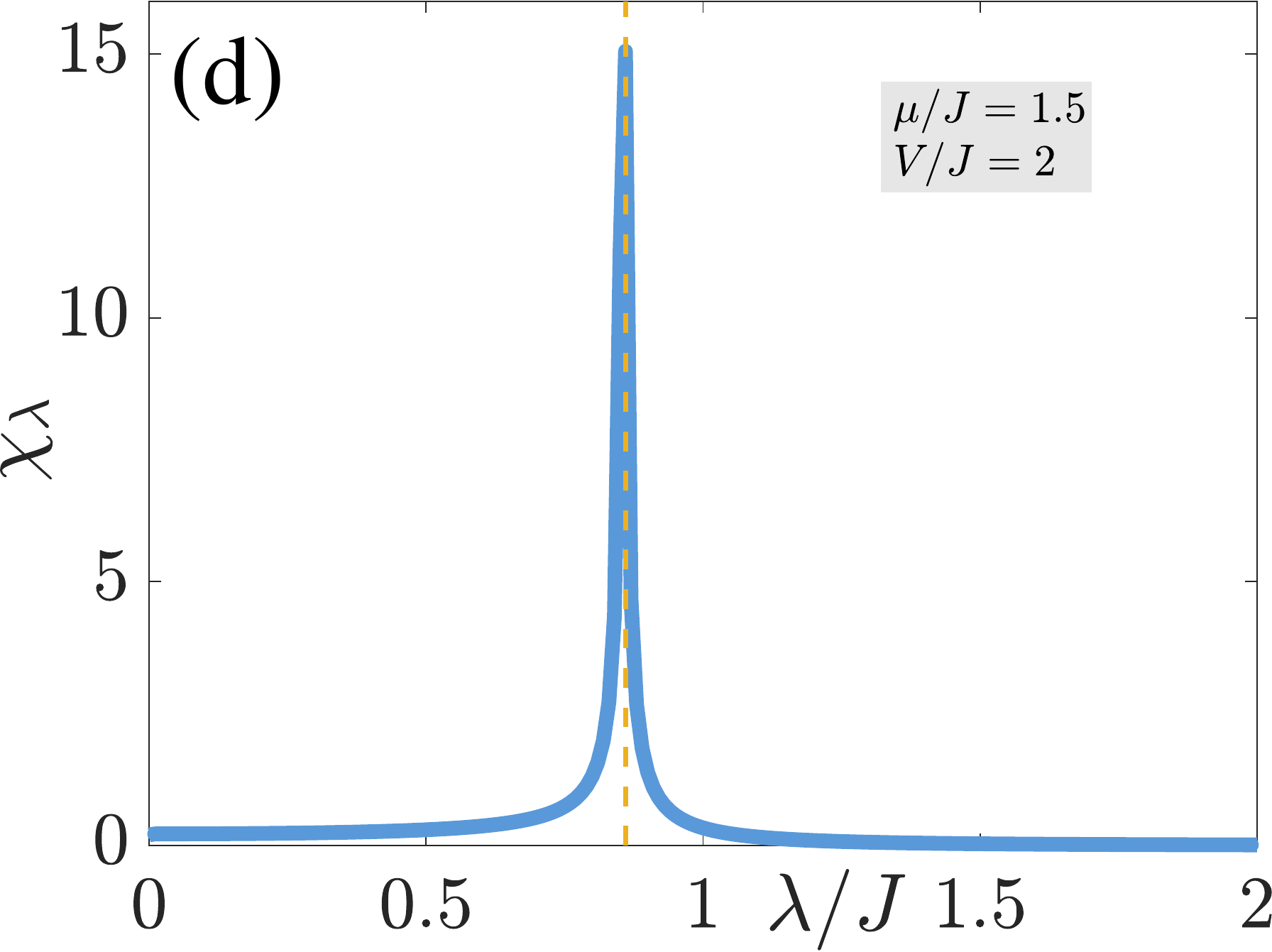}
	\caption{Fidelity susceptibility $\chi_\lambda$ along $\lambda$, at cuts through the phase diagram. 
		(a) Fixed $\lambda=0.25$, 
		(b) fixed $\lambda=1.25$, 
		(c) fixed $\mu=0.25$, 
		(d) fixed $\mu=1.5$. 
		Strong divergencies indicate the quantum phase transitions. 
		Since the phase transition line emanating from the ideal Colmen's phase transition at $\lambda=0$, $\mu/J\approx0.655$ runs close to parallel to $\lambda$, the corresponding $\chi_\lambda$ in panel (a) shows only a small, but nevertheless clearly visible divergence. Note the much stronger divergence for $\chi_\mu$, displayed in Fig.~\ref{fig:FidSusmu}a.  
		In all data sets $V=2$.}
	\label{fig:FidSuslam}
\end{figure}

\begin{figure}[ht!]
	\centering
	\includegraphics[width=.23\textwidth]{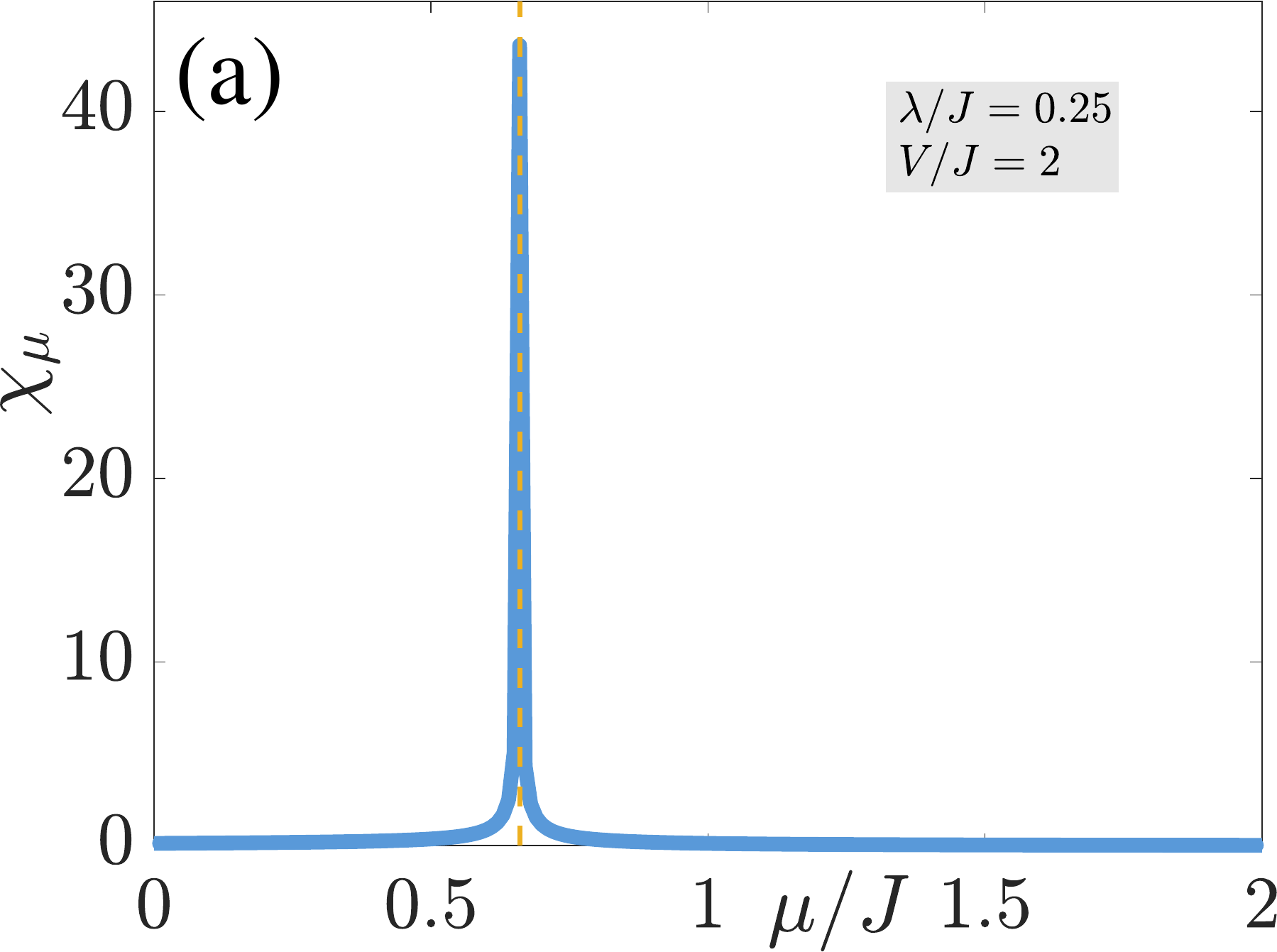}\quad
	\includegraphics[width=.23\textwidth]{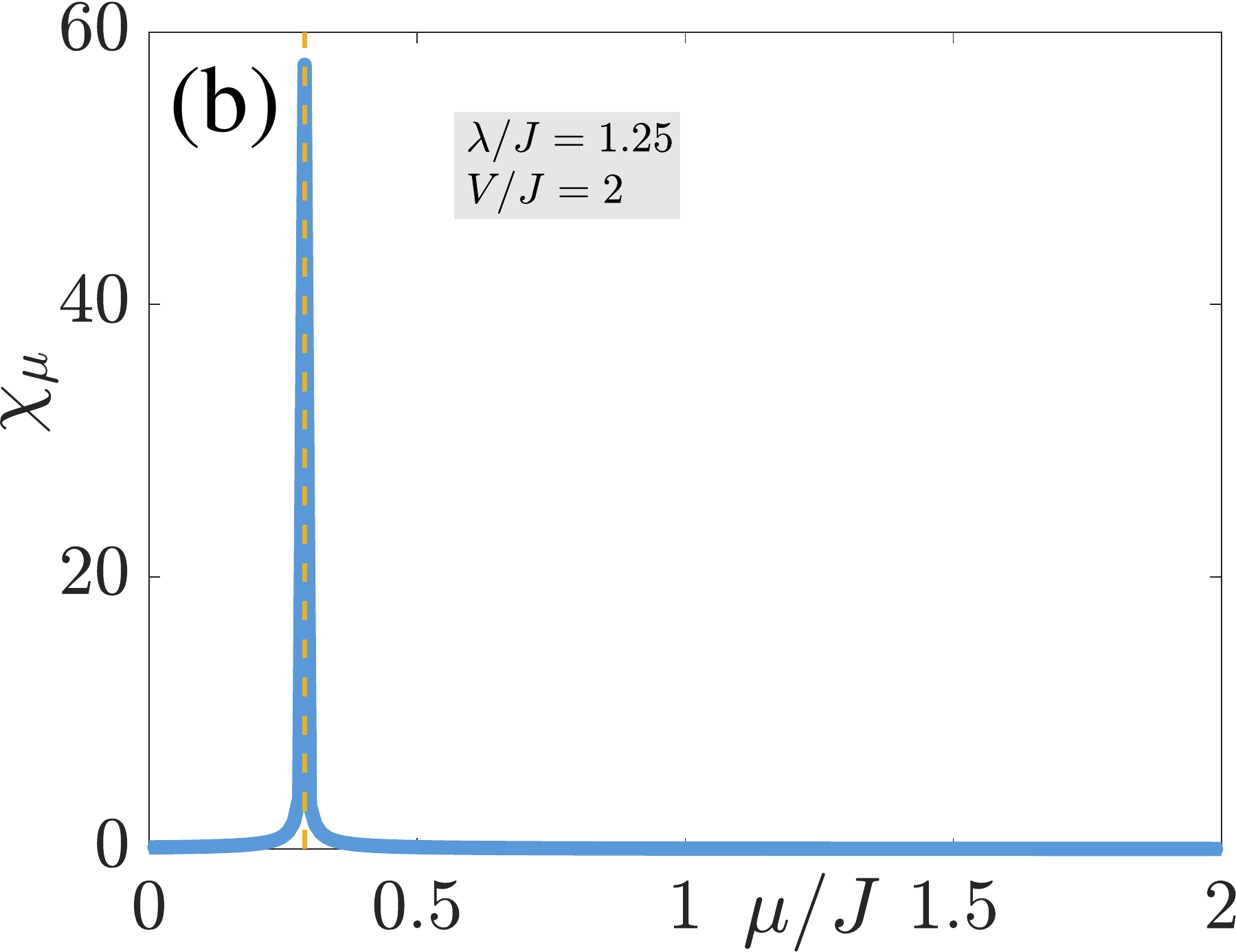}\\
	\includegraphics[width=.23\textwidth]{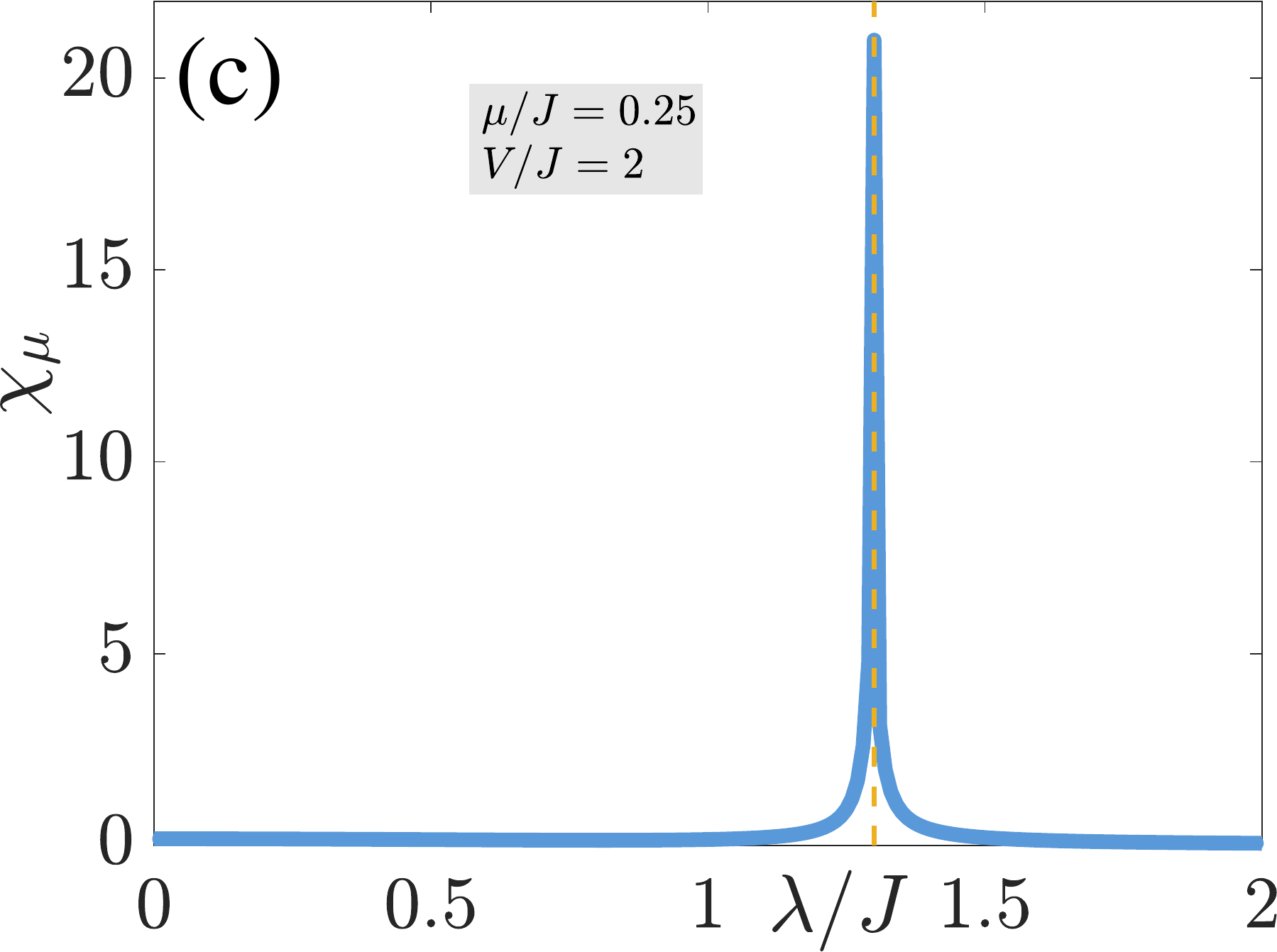}\quad
	\includegraphics[width=.23\textwidth]{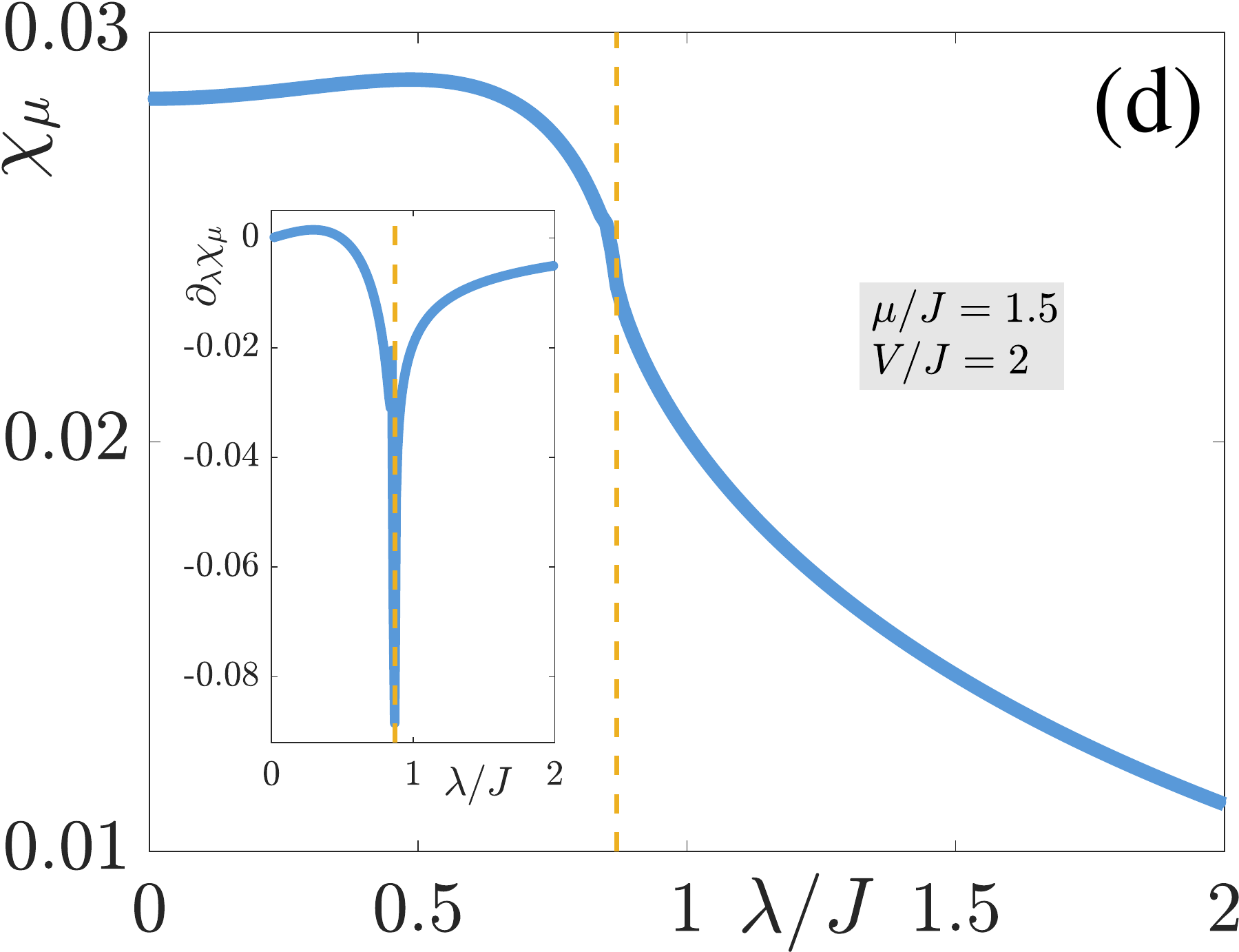}
	\caption{
		As Fig.~\ref{fig:FidSuslam}, but for $\chi_\mu$. 
		Strong divergencies in panels (a-c) indicate the quantum phase transitions. 
		The transition line probed in panel (d) lies almost parallel to $\mu$, such that no divergence is seen in $\chi_\mu$. Nevertheless, its derivative shows a clear nonanalytic behavior. 
		Note also the strong divergence for $\chi_\lambda$, displayed in Fig.~\ref{fig:FidSuslam}d.
		}
	\label{fig:FidSusmu}
\end{figure}

Figures~\ref{fig:FidSuslam} and \ref{fig:FidSusmu} present the fidelity susceptibility across cuts through parameter space. The sharp divergencies permit one to pinpoint the quantum phase transition seen in Fig.~\ref{fig:PD} of the main text.

\section{Renormalized gauge symmetry of physical subspace}

In this section, we derive the renormalized symmetry $\tilde{H}_G$ that governs the ground-state phase diagram at low error strength, starting from an effective Hamiltonian $\tilde{H}$ that almost commutes with $H_G$. We first make rigorous statements based on a `gauge protection theorem' \cite{Abanin2017,Halimeh2020e}  that show the existence of a renormalized gauge theory at least to exponential accuracy in $\lambda/V$. These are then followed by an analysis based on a perturbative Schrieffer--Wolff transformation, which we use to construct $\tilde{H}_G$ to low order. 

The considered microscopic theory reads $H=VH_G + H_0 + \lambda H_1$. 
Here, $H_0$ commutes with $H_G$, while we assume $H_1$ to contain only terms that couple states with different eigenvalues with respect to $H_G$. 
The target gauge-invariant sector is the so-called physical subspace of $G_j\ket{\psi}_\mathrm{phys}=0$, $\forall j$, which coincides with the zero-eigenvalue subspace of the energy protection term $H_G$, $H_G\ket{\psi}_\mathrm{phys}=0$. 
We denote the projector on the physical subspace as $\mathcal{P}$ and define $\mathcal{Q}=\mathbb{I}-\mathcal{P}$ as the projector on all gauge-violating states. 
In what follows, we assume $V$ to be the dominant energy scale of the problem, and formally set $H_0=\mathcal{O}(\lambda)$. 

\subsubsection{Gauge protection theorem}
We start by making rigorous statements based on a recently introduced gauge protection theorem \cite{Halimeh2020e}, which applies results from Ref.~\cite{Abanin2017}. 

To this end, we first repeat a few definitions from Refs.~\cite{Abanin2017,Halimeh2020e}. 
We define $\Lambda$ as a finite subset of the $d$-dimensional spatial cubic lattice $\mathbb{Z}^d$ on which the gauge theory lives. 
Further, we the algebra of bounded operators acting on the total Hilbert space $\mathcal{H}_\Lambda$, equipped with the standard operator norm, is $\mathcal{B}_\Lambda$. 
The operators of the form $O_S\otimes\mathcal{I}_{\Lambda \backslash S}$ with $S\subset \Lambda$ form the subalgebra $\mathcal{B}_S \subset \mathcal{B}_\Lambda$. 
Any operator $X$ can be decomposed as $X = \sum_{S\in \mathcal{P}_c(\Lambda)} X_S$, where $X_S \in \mathcal{B}_S$ is a so-called `potential'. Moreover, $\mathcal{P}_c(\Lambda)$ denotes the set of finite, connected subsets of $\Lambda$. 
One can define a family of norms on potentials as  
\begin{equation}
||X||_\kappa = \sup\limits_{j\in\Lambda} \sum_{S\in \mathcal{P}_c(\Lambda):S\ni j} e^{\kappa |S|} ||X_S||.
\end{equation}
These norms are parametrized by a rate $\kappa >0$ that gives different weights to operators with different spatial support. 
The supremum selects the lattice site $j$ with the largest sum of weighted norms of the operators $X_S$ that have support on $x$.

Using these definitions, we can derive the following \cite{Abanin2017,Halimeh2020e}. 
Given a protection Hamiltonian such as $VH_G$ with integer spectrum. Assume there exists a $\kappa_0$ that permits for defining the relevant energy scale as $V_0 = \frac{54\pi}{\kappa_0^2}(||H_{0}||_{\kappa_0}+2||H_{1}||_{\kappa_0})$, and assume the following two conditions to hold, 
\begin{subequations}
	\begin{align}
	&V \ge \frac{9\pi||H_{1}||_{\kappa_0}}{\kappa_0},\\
	\label{eq:nstar}
	&n_* = \lfloor \frac{V/V_0}{(1+\ln{V/V_0})^3}\rfloor - 2 \ge 1. 
	\end{align}
\end{subequations}

Then, there exists a unitary operator $Y$ such that 
\begin{align}\nonumber
H^\prime= &YHY^\dag =VH_G + {H}^\prime_0 + {H}^\prime_1,
\end{align}
with $[{H}^\prime_0,H_G]=0$ and $H_1^\prime$ containing all residual terms that couple different eigenvalues of $H_G$. 
Further
\begin{align}
||{H}^\prime_{0}-H_{0}||_{\kappa_{n_*}}&\le C(V_0/V), \\
||{H}^\prime_{1}||_{\kappa_{n_*}}&\le (2/3)^{n_*}||H_{1}||_{\kappa_0},
\end{align}
where $\kappa_{n_*}=\kappa_0[1+\log(1+n_*)]^{-1}$ and $C$ is a constant. 
Thus, we have obtained a renormalized gauge theory ${H}^\prime_{0}$ that is perturbatively close in $V_0/V$ to $H_{0}$. 
Even more importantly, the new contribution that fails to commute with $H_G$, ${H}^\prime_{\rm 1}$, is exponentially small in $V/V_0$, 
\begin{align}
||[H_G,{H}^\prime]||_{\kappa_{n_*}}&\le (2/3)^{n_*}||H_{1}||_{\kappa_0}\,.
\end{align}

By swapping the unitary transformation from ${H}^\prime$ to $H_G$, we can exploit this last relation to define a renormalized approximate symmetry ${H}^\prime_G = Y^\dag H_G Y$, which fails to commute with $H$ at the most with an exponentially small contribution, independent of system size, 
\begin{align}
||[{H}^\prime_G,H]||_{\kappa_{n_*}}&\le (2/3)^{n_*}||H_{1}||_{\kappa_0}\,. 
\end{align}
As is often the case for such rigorous mathematical proves, the involved bounds are often not tight and the relevant constants can be very large (see, e.g., \cite{Halimeh2020e} for explicit numbers). Nevertheless, conceptually this result is extremely important, as it shows us that a gauge symmetry in the physical subspace can be obtained to exponential accuracy even in the thermodynamic limit. Numerics then demonstrates that (i) the required gauge protection strength is actually very moderate and (ii) the exponentially small upper bound does not mean the system is immediately driven into a gauge-violating phase.

\subsubsection{Schrieffer--Wolff transformation}

To derive the form of the approximate symmetry to low order in $\lambda/V$, we now turn to a Schrieffer--Wolff transformation \cite{Bravyi2011}. The convergence of such a perturbative method in a quantum many-body system is not necessarily a given, as the operator norm of the perturbation rises above all bounds as $||\lambda H_1||\propto N$. Nevertheless, given a dominant energy scale (here $V$), its low-order results give a powerful approximation to construct an effective Hamiltonian $\tilde{H}$ restricted to a low-energy subspace (here $\mathcal{P}$), as is also corroborated by our numerical results. 

The Schrieffer--Wolff transformation, $\tilde{H}=\rme^S H \rme^{-S}$, is generated by an anti-hermitian operator $S=\sum_k S_k$. It can be constructed as an expansion in powers of $\lambda/V$, $S_k=\mathcal{O}(\lambda/V)^k$, such that contributions leaving $\mathcal{P}$ are cancelled order by order. 
Reference~\cite{Bravyi2011} derives all orders $S_k$. Here, we will explicit need only 
\begin{align}
	S_1&=\mathcal{L}(O)\,,\\
	S_2&=-\mathcal{L}([D,S_1])\,,
\end{align}
where we have introduced the diagonal perturbation $D=\mathcal{P}H_0\mathcal{P} + \mathcal{Q}(H_0+\lambda H_1)\mathcal{Q}$ and the off-diagonal perturbation $O=\mathcal{P}\lambda H_1\mathcal{Q} + \mathcal{Q}\lambda H_1\mathcal{P}$ (exploiting $[H_0,H_G]=0$ and that $\mathcal{P}H_1\mathcal{P}=0$). 
Further, we defined the superoperator
\begin{align}
\mathcal{L}(X)&=\sum_{n,m}\frac{\ket{n}\bra{n}X\ket{m}\bra{m}}{E_n-E_m}\,,
\end{align}
where $\ket{n,m}$ are eigenstates of the unperturbed Hamiltonian $VH_G$ with eigenenergies $E_{n,m}$. Though we will not need it in the following, for completeness we nevertheless give the form of $\tilde{H}$. 
Define $H_1=\sum_j h_{1,j}$, where $h_{1,j}$ are local operators that break gauge symmetry. Then, we obtain 
\begin{subequations}
	\label{eq:tildeH}
	\begin{align}
\mathcal{P}\tilde{H}\mathcal{P}&=H_0 \mathcal{P} - \lambda^2 \mathcal{P} H_1 \mathcal{Q} \sum_{m\in \mathcal{Q}}\frac{\ket{m}\bra{m}}{E_m} \mathcal{Q} H_1 \mathcal{P}+\mathcal{O}(\frac{\lambda^3}{V^2})\,,\\
&=H_0 \mathcal{P} - \frac{\lambda^2}{2V} \mathcal{P} \sum_j h_{1,j}^2 \mathcal{P}+\mathcal{O}(\frac{\lambda^3}{V^2})\,.
\end{align}
\end{subequations}
Here, we used $H_G\ket{\psi}_\mathrm{phys}=0$, $H_G\ket{m}=2V\ket{m}$ for all  $\ket{m}$ that can be accessed from $\mathcal{P}$ by a single application of $H_1$, and $\mathcal{P} H_1\mathcal{Q} H_1 \mathcal{P}=\mathcal{P} \sum_j h_{1,j}^2 \mathcal{P}$.

If we truncate the Schrieffer--Wolff transformation at a given $\mathcal{O}((\lambda/V)^K)$, then within the ground space (the physical subspace) the effective theory $\tilde{H}$ and $H_G$ approximately commute as
\begin{align}
||\mathcal{P}\left[H_G,\frac{\tilde{H}}{V}\right]\mathcal{P}||&\leq \mathcal{O}((\frac{\lambda}{V})^{K+1})\,, 
\end{align}
where $||\bullet||$ denotes the operator norm. 
We can use this expression to rewrite 
\begin{align}
||\tilde{\mathcal{P}}\left[\tilde{H}_G,\frac{H}{V}\right]\tilde{\mathcal{P}}||&\leq \mathcal{O}((\frac{\lambda}{V})^{K+1})\,, 
\end{align}
where we defined $\tilde{\mathcal{P}}=\rme^{-S} \mathcal{P}\rme^{S}$ and $\tilde{H}_G=\rme^{-S} H_G \rme^{S}$. 
Moreover, 
\begin{align}
||\tilde{\mathcal{P}}\tilde{H}_G\tilde{\mathcal{P}}||&=||{\mathcal{P}}{H}_G{\mathcal{P}}||=0 
\end{align}
In other words, there exists a perturbatively renormalized symmetry generator $\tilde{H}_G$, which approximately commutes with $H$ and which defines a renormalized physical subspace $\tilde{\mathcal{P}}$. 
$\tilde{\mathcal{P}}$ contains the ground state $\ket{\psi_0}$ of $H$, so that we will get $\bra{\psi_0}\tilde{H}_G\ket{\psi_0}=0$, as long as we remain in the gauge-symmetry-retaining phases and the perturbative Schrieffer--Wolff transformation converges. 

An explicit, low-order approximation to $\tilde{H}_G$ can be derived proceeding analogously to the derivation of Eq.~\eqref{eq:tildeH}. Using $\tilde{\mathcal{P}}={\mathcal{P}}+\mathcal{O}({\lambda}/{V})$, we obtain 
\begin{subequations}
\begin{align}
\tilde{\mathcal{P}}\tilde{H}_G\tilde{\mathcal{P}} &= \tilde{\mathcal{P}}\big[{H}_G + \frac{\lambda}{V}H_1 \\ 
& \qquad\,\,\,\,\, + \frac{\lambda^2}{V}H_1\mathcal{Q}\sum_{m\in \mathcal{Q}}\frac{\ket{m}\bra{m}}{E_m}\mathcal{Q}H_1\big]\tilde{\mathcal{P}}+\mathcal{O}((\frac{\lambda}{V})^3) \nonumber\\
&= \tilde{\mathcal{P}}\left[{H}_G + \frac{\lambda}{V}H_1 + \frac{1}{2}\frac{\lambda^2}{V^2}\sum_j h_{1,j}^2 \right]\tilde{\mathcal{P}}+\mathcal{O}((\frac{\lambda}{V})^3)\,.
\end{align}
\end{subequations}

The ground-state expectation value of $\tilde{H}_G$ up to this order is plotted in Fig.~\ref{fig:Gausslaw}. Its value is much suprressed as compared to the expectation value of the bare ${H}_G$, showing the emergence of a renormalized symmetry that is well approximated to this order. 
Importantly, for a typical local error term, this is a local few-body observable that can be accessed in realistic experiment.

Note that, even though ${H}_G$ is a global symmetry operator, within the physical subspace it behaves exactly as the generators of Gauss's law, as ${H}_G\ket{\psi}_\mathrm{phys}=0$ can be fulfilled if and only if $G_j\ket{\psi}_\mathrm{phys}=0$, $\forall j$. The same holds for $\tilde{H}_G$, which defines renormalized Gauss's law generators $\tilde{G}_j=\rme^{-S} G_j \rme^{S}$ with $\tilde{\mathcal{P}}\tilde{G}_j\tilde{\mathcal{P}}=0$. Thus, the addition of $V{H}_G$ to the Hamiltonian restores a renormalized gauge symmetry for the ground state in the (dressed) physical subspace. One can extend this procedure to other desired gauge sectors, just be redefining $H_G=\sum_j (G_j-g_j)^2$ to vanish for other target eigenvalues $g_j$ of $G_j$.

\subsection{Gutzwiller-type mean-field treatment}

In order to obtain a better insight into the observed phase diagram, it is instructive to consider a Gutzwiller-type mean-field decomposition between matter (m) and gauge (g) fields, where we assume the ground state to separate as $\ket{\psi_0}=\ket{\psi_0}_\mathrm{g}\otimes \ket{\psi_0}_\mathrm{m}$. 
For small $\lambda$, we expect this formulation to be not very quantitative, since in a gauge theory matter and gauge fields are intrinsically tied to each other. However, it can give us some qualitative insight on the phase transition occurring at large $\lambda$, which lifts the intrinsic coupling between matter and gauge fields.

\subsubsection{Gauge fields: Ising model in transverse and longitudinal field and transition at large $\mu$}

We first study the gauge fields in this mean-field decomposition, replacing the matter fields by their expectation values. We obtain  
\begin{align}\label{eq:HamMFs}
H^\mathrm{GW,g}=H_0^\mathrm{GW,g}+\lambda H_1^\mathrm{GW,g} + V H_G^\mathrm{GW,g}\,,
\end{align}
with
\begin{align}
H_0^\mathrm{GW,g}&=-J\sum_j\big(s^+_{j,j+1} \braket{\sigma^-_j \sigma^-_{j+1}} + \text{H.c.}\big),\\
H_1^\mathrm{GW,g}&=2\sum_j s^x_{j,j+1}\\
H_G^\mathrm{GW,g} &= \sum_j (G_j^\mathrm{GW,g})^2,\\
&=2\sum_j s^z_{j-1,j}s^z_{j,j+1}+\sum_j s^z_{j,j+1}(2+\braket{\sigma_j^z}+\braket{\sigma_{j+1}^z})\,,\nonumber
\end{align}
where we used the mean-field version of the Gauss's law generator
\begin{align}
G_j^\mathrm{GW,g}=\frac{(-1)^j}{2}\big[2\big(s^z_{j,j+1}+s^z_{j-1,j}\big)+1+\braket{\sigma^z_j}\big].
\end{align}

Thus, in this approximation, the gauge fiels are governed by the antiferromagnetic Ising model in transverse and longitudinal field
\begin{subequations}
\begin{align}\nonumber
H^{\text{GW},g}=
&\,2V\sum_js^z_{j-1,j}s^z_{j,j+1}+\gamma_z\sum_js^z_{j,j+1}\\
&+\gamma_y\sum_js^y_{j,j+1}+\gamma_x\sum_js^x_{j,j+1},\\
\gamma_x = &\,2\lambda-J(\langle\sigma^-_j\sigma^-_{j+1}\rangle+\langle\sigma^+_j\sigma^+_{j+1}\rangle),\\
\gamma_y =&\, -iJ\big(\langle\sigma^-_j\sigma^-_{j+1}\rangle-\langle\sigma^+_j\sigma^+_{j+1}\rangle\big),\\
\gamma_z =&\,2V(1+\frac{\langle\sigma^z_j\rangle+\langle\sigma^z_{j+1}\rangle}{2}).
\end{align}
\end{subequations}

In the limit $\mu\to\infty$, matter particles are infinitely heavy and we have $\braket{\sigma_j^z}=-1$ as well as $\langle\sigma^-_j\sigma^-_{j+1}\rangle=0$, and thus obtain a vanishing longitudinal field $\gamma_z=0$. 
The transverse field $\gamma_x$ will drive a quantum phase transition between an Ising antiferromagnet and a paramagnetic phase, which---in this limit---will occur at $\gamma_x=V$, i.e., $\lambda=V/2$. 

At $\mu<\infty$, the longitudinal field enters the game. Notably, translational invariance by two sites ensures $\braket{\sigma_j^z}+\braket{\sigma_{j+1}^z}=\mathrm{const}$, independent of $j$, making the longitudinal field $\gamma_z$ homogeneous. 
In a ferromagnet, such a homogeneous transverse field would negate entering the ordered phase at arbitrarily small strength $\gamma_z\neq 0$. 
For antiferromagnetic interactions, however, a sufficiently large strength $\gamma_z>0$ is required to invalidates the symmetry-breaking transition \cite{PhysRevE.99.012122}.
Conversely, for $\gamma_z>0$, a $\gamma_x<V$ is sufficient to enter the paramagnetic phase. 
Thus, for $\mu<\infty$, we expect the system to undergo a phase transition at some value of $\lambda\leq V/2$, which is indeed what we observe as an almost horizontal line in Fig.~\ref{fig:PD} of the main text. 

At small $\mu$, $\braket{\sigma_j^z}+\braket{\sigma_{j+1}^z}$ increases and $\gamma_z$ can become large. Thus, the gauge-field mean-field Hamiltonian cannot easily explain the observed phase transition at small $\mu$. Its origin becomes clearer through the matter fields.

\subsubsection{Matter fields: XY spin chain in transverse field and transition at small $\mu$}

Replacing the gauge fields with their expectation value, we obtain a Gutzwiller-type mean-field Hamiltonian for the matter fields  
\begin{align}\label{eq:HamMFsigma}
H^\mathrm{GW,m}=H_0^\mathrm{GW,m}+\lambda H_1^\mathrm{GW,m} + V H_G^\mathrm{GW,m}\,,
\end{align}
with
\begin{align}
H_0^\mathrm{GW,m}&=-J\sum_j\big(\braket{s^+_{j,j+1}} \sigma^-_j \sigma^-_{j+1} + \text{H.c.}\big)+\mu\sum_j\sigma^+_j \sigma^-_j,\\
H_1^\mathrm{GW,m}&=\sum_j(\sigma^-_j \sigma^-_{j+1}+\text{H.c.}) \\
H_G^\mathrm{GW,m} &= \sum_j (G_j^\mathrm{GW,m})^2,\\
&=\sum_j\frac{1}{2}\sigma^z_j\big[2\big(\braket{s^z_{j,j+1}}+\braket{s^z_{j-1,j}}\big)+1\big]\,,\nonumber
\end{align}
where we used the mean-field version of the Gauss's law generator
\begin{align}
G_j^\mathrm{GW,m}=\frac{(-1)^j}{2}\big[2\big(\braket{s^z_{j,j+1}}+\braket{s^z_{j-1,j}}\big)+1+\sigma^z_j\big].
\end{align}

Collecting all terms, we obtain 
\begin{align}
H^\mathrm{GW,m}&=\sum_j\big(-J_j^\mathrm{eff} \sigma^-_j \sigma^-_{j+1} + \text{H.c.}\big)+\delta_z \sum_j\sigma^z_j ,
\end{align}
with $J_j^\mathrm{eff}=-J \braket{s^+_{j,j+1}}+\lambda$ and 
$\delta_z=\frac{\mu}{2}+\frac{V}{2}\big[2\big(\braket{s^z_{j,j+1}}+\braket{s^z_{j-1,j}}\big)+1\big]$ (again, due to translational symmetry by two matter sites, $\delta_z$ is independent of $j$). 
After a staggered $\pi$ rotation of the matter fields $\sigma$ around the $y$-axis, this Hamiltonian takes the more familiar form of an XY spin model in a staggered transverse field, $\sum_j\big(-J_j^\mathrm{eff} \sigma^+_j \sigma^-_{j+1} + \text{H.c.}\big)+\delta_z \sum_j(-1)^j\sigma^z_j$. 
In the absence of a homogeneous transverse field, and assuming $J_j^\mathrm{eff}$ constant, this model undergoes a phase transition at $\delta_z=0$ \cite{dutta2015quantum}. 
In the limit of $\lambda\to\infty$, we get $J_j^\mathrm{eff}\approx \lambda$ and, since $\braket{s^z_{j,j+1}}\approx 0$, $\delta_z\approx  \frac{\mu}{2}+\frac{V}{2}$. Thus, we expect a transition at $\mu=-V$, which is indeed what we observe numerically. 

When decreasing $\lambda$, this transition line driven by the matter field approaches the line that is smoothly connected to $\mu=\infty$, $\lambda=V/2$, which is driven by the gauge field. In that region, the entire system becomes strongly correlated, and the two Gutzwiller mean-field transitions do not cross but are inflected. This gives the phase structure we observe in Fig.~\ref{fig:PD} of the main text: one line that smoothly connects from the ideal Coleman's phase transition at $\lambda=0$, $\mu/J\approx 0.655$ to the gauge-field driven transition point at $\lambda=\infty$, $\mu=-2V$, and another line that smoothly connects from the matter-field driven transition point $\mu=\infty$, $\lambda=V/2$ to the gauge-field driven transition point.

\begin{figure}[ht!]
	\centering
	\includegraphics[width=.23\textwidth]{{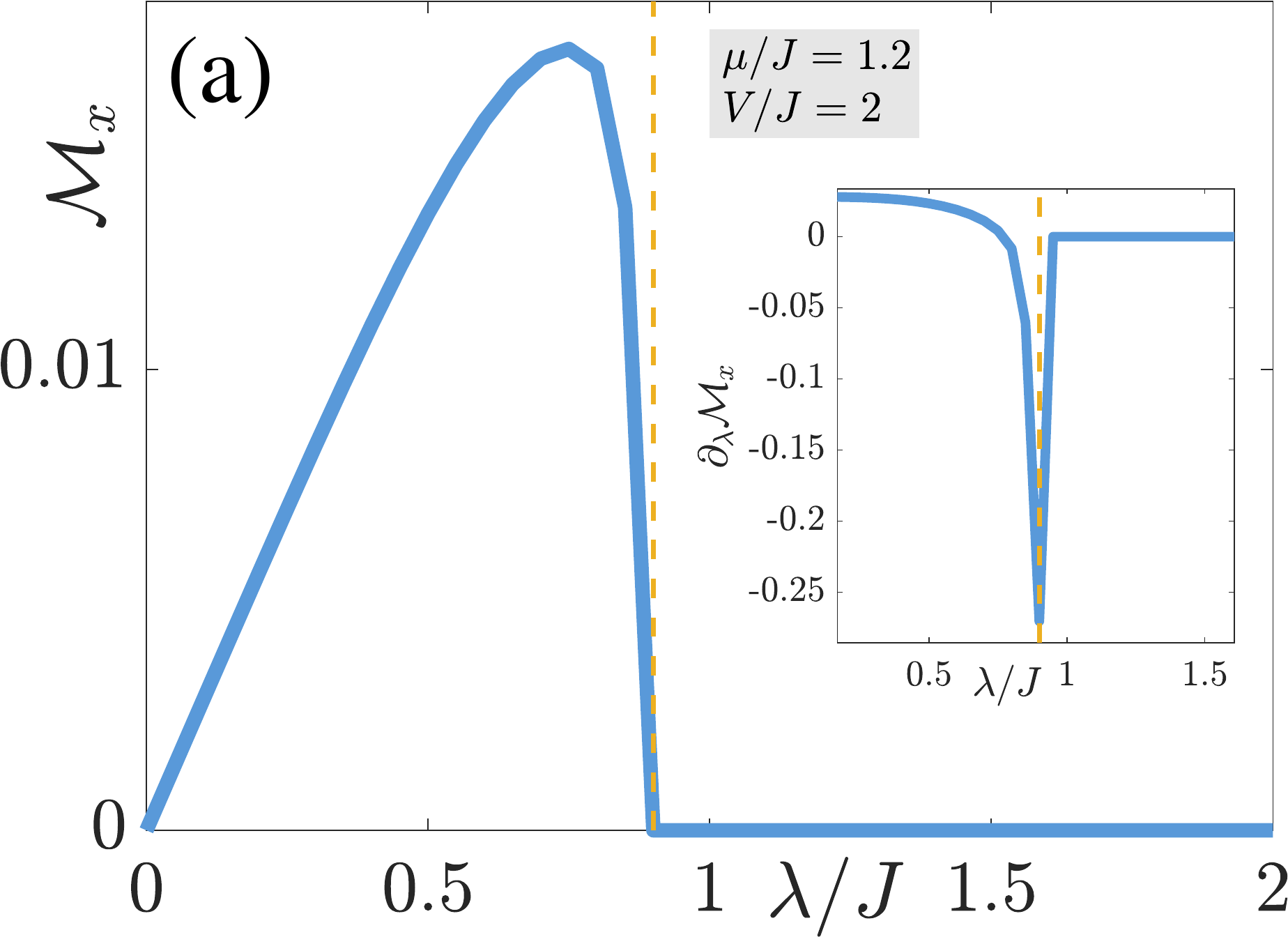}}\quad\includegraphics[width=.23\textwidth]{{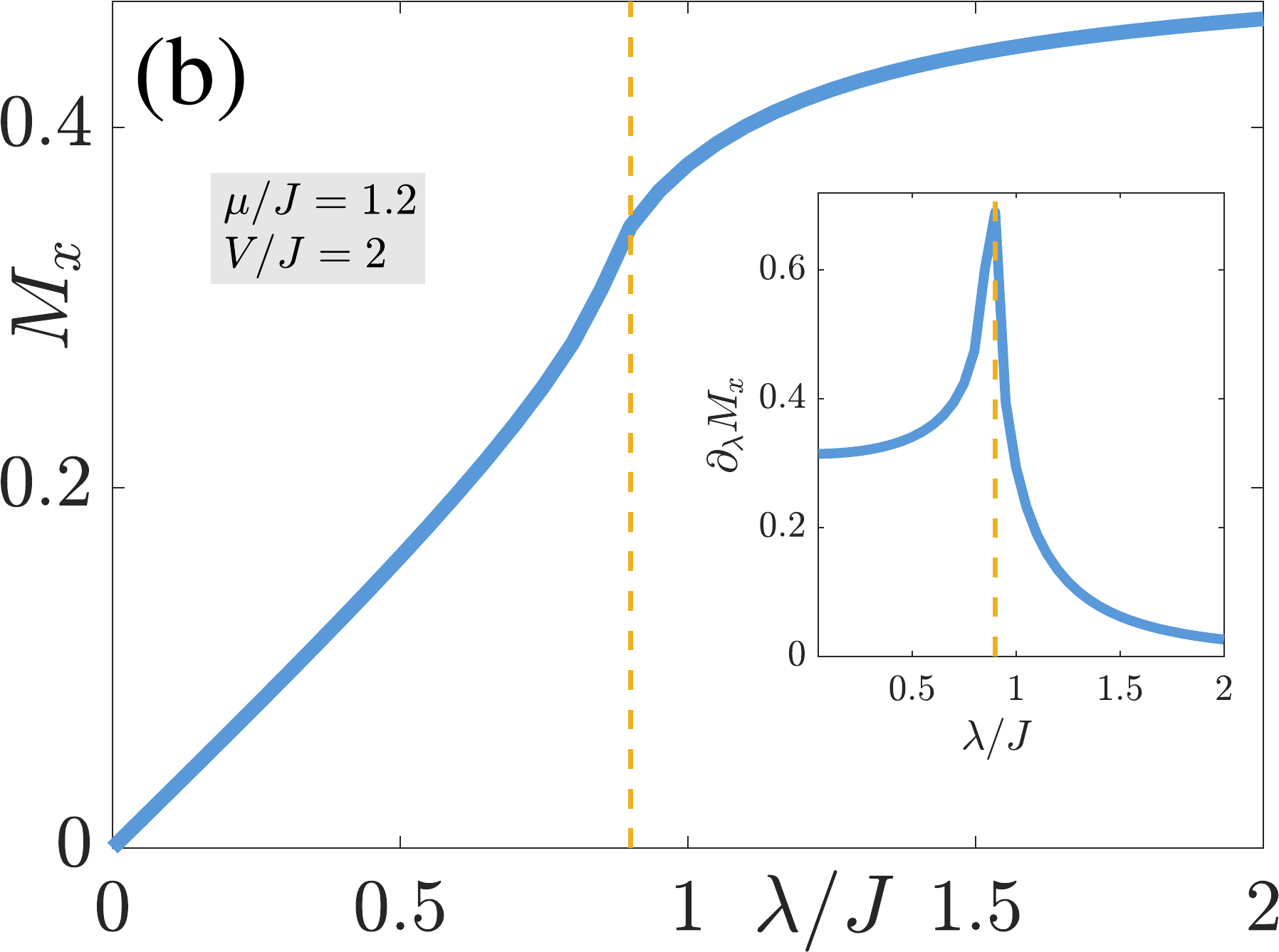}}\\
	\includegraphics[width=.23\textwidth]{{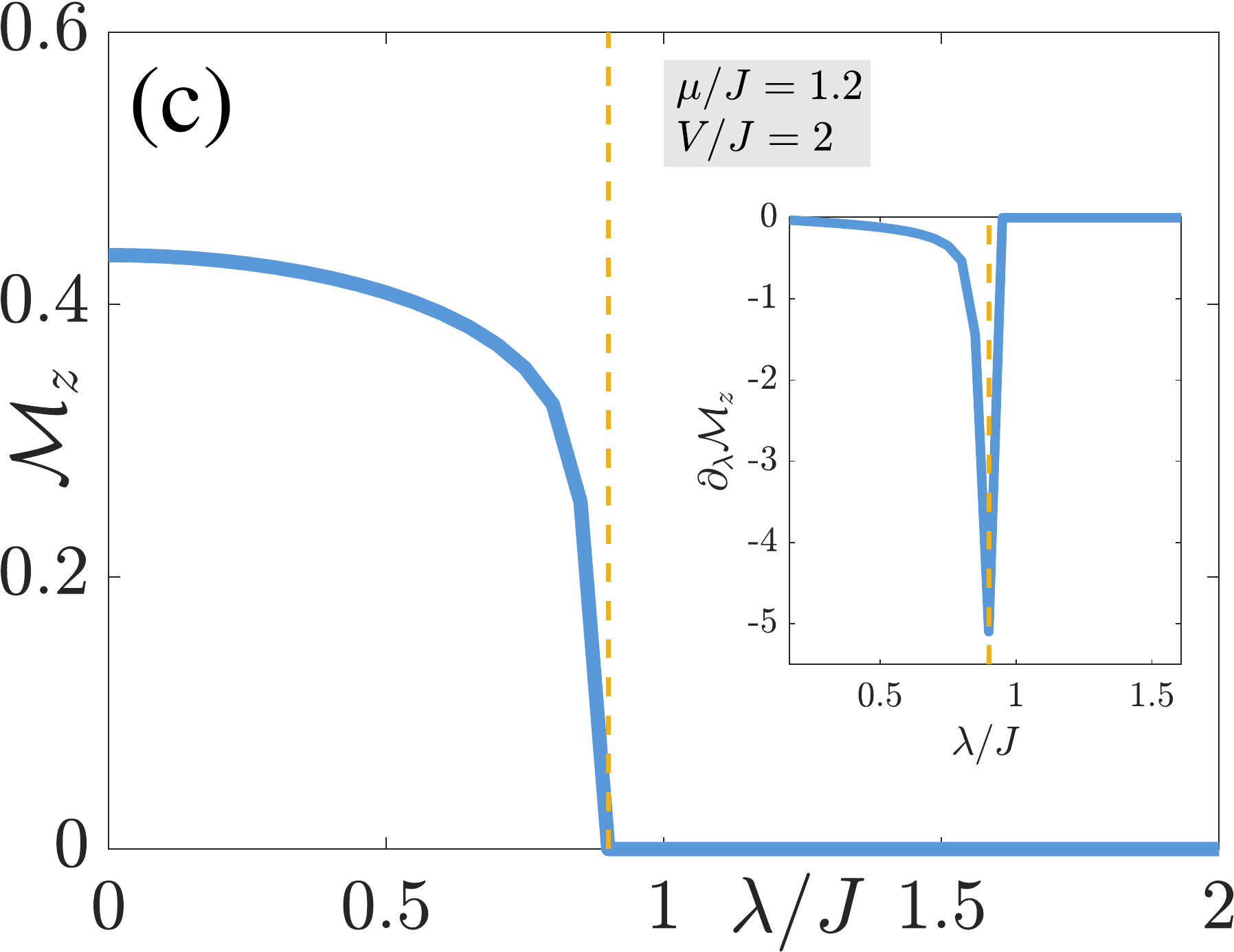}}\quad\includegraphics[width=.23\textwidth]{{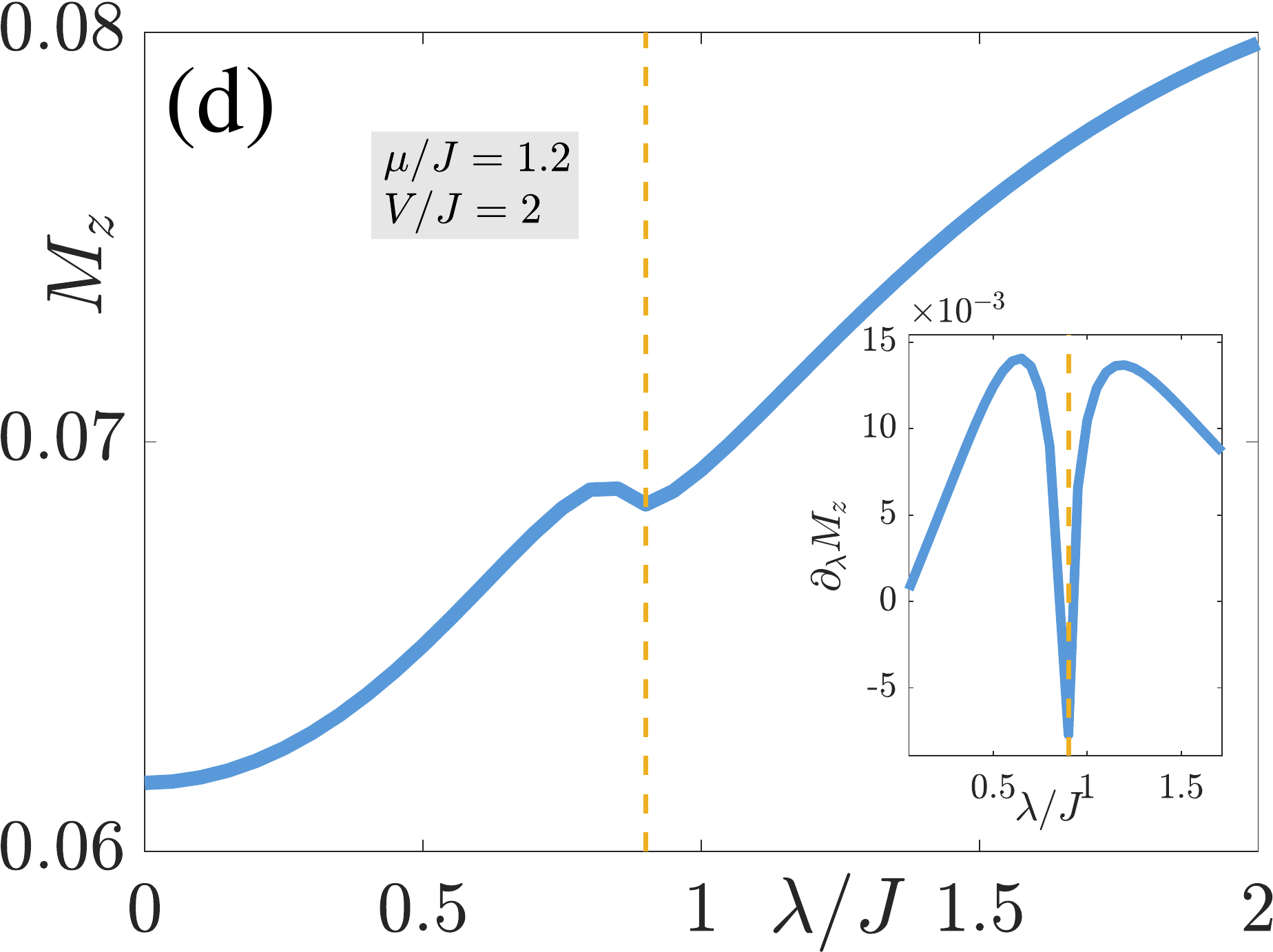}}
	\caption{
		Cut along $\lambda$ at fixed $\mu=1.2J$, $V=2J$, for the local observables 
		$\mathcal{M}_{x,z}=(\braket{s^{x,z}_{2n-1,2n}}-\braket{s^{x,z}_{2n,2n+1}})/2$ and 
		$M_{x,z}=(\braket{s^{x,z}_{2n-1,2n}}+\braket{s^{x,z}_{2n,2n+1}})/2$ (the system has translational invariance by 2 sites). 
		(a) At nonzero $\lambda$, the staggered magnetization $\mathcal{M}_{x}$ becomes nonzero, indicating a Peierls-type physics. 
		(b) The homogeneous magnetization ${M}_{x}$ increases approximately linearly at low $\lambda$, which is reflected in a contribution quadratic in $\lambda/V$ to the dressed gauge violation defined by Eq.~\eqref{eq:tildeH}. 
		(c) The order parameter of Coleman's phase transition, the staggered magnetization $\mathcal{M}_{z}$ remains nonzero for a large parameter window, up to a quantum phase transition at around $\lambda/J\approx 0.9$. 
		(d) The homogeneous magnetization $M_{z}$ increases with $\lambda$. 
		At the quantum phase transition, all these observables show nonanalyticities. 
	}
	\label{fig:Peierls}
\end{figure}

In the above analysis, we assumed that $J_j^\mathrm{eff}$ is homogeneous, i.e., independent of $j$. That is true in the region that contains the mean-field transition line connecting from the ideal Coleman's phase transition to the limit $\lambda=\infty$. However, $J_j^\mathrm{eff}$ can also become dimerized due to a non-zero staggered magnetization along the $x$ direction. 
This is illustrated in Fig~\ref{fig:Peierls}, which shows a cut along $\lambda$ for fixed $\mu=1.2J$. 
Such Peierls phase transitions have been discussed in similar models that do not host a gauge symmetry \cite{Cuadra2018}. Indeed, the loss of the bare, ideal gauge symmetry enables this physics to take place also here.

\subsection{Coupled self consistent mean field results}

One can obtain mean-field estimates for the phase transitions across the entire phase diagram by solving the coupled mean-field model given by Eqs.~\eqref{eq:HamMFsigma} and~\eqref{eq:HamMFs} in a self-consistent way.

As discussed above, at $\lambda = \infty$, we have a phase transition in $H^\mathrm{GW,m}$ at $\delta_z = 0$. 
This remains true for smaller values of $\lambda$ as long as we assume there is no staggering of the gauge fields in the $x$ direction (which is justified a posteriori, since the resulting mean-field prediction for the phase transition lies indeed in a region where the staggered magnetization in $x$ direction vanishes). 
At the point where $\delta_z = 0$, we find numerically 
\begin{subequations}
	\begin{align}\nonumber
	\langle\sigma^-_j\sigma^-_{j+1}\rangle = 
	& -0.3183\,, \\
	\nonumber \langle\sigma^z_j\rangle = 
	&  0.0\,,
	\end{align}
\end{subequations}
independent of $V,\lambda,\mu$. 

We can plug this into $H^\mathrm{GW,g}$ and numerically evaluate the expectation values of $\langle s^z_{j,j+1} \rangle$ at different $\lambda,V$. Together with $\delta_z = 0$ this fixes uniquely the critical $\mu,\lambda,V$, which gives the transition line connecting smoothly to the point $\lambda=\infty$, $\mu=-V$, as plotted in Fig.~\ref{fig:PD}.

In addition, at large $\mu$ we obtain an antiferromagnetic Ising model in a staggered transverse and homogeneous longitudinal magnetic field. Numerically, we find a critical point at approximately 
\begin{equation}
\label{eq:meanfield_elips}
4\gamma_x^2 + \gamma_z^2 = 4 V^2\,.
\end{equation}
For every $\mu$ and $V$, we set $\langle s^z_{j,j+1} \rangle,\langle s^+_{j,j+1} \rangle, \langle\sigma^-_j\sigma^-_{j+1}\rangle, \langle\sigma^z_j\rangle$ to their initial values at $\lambda=0$. We then iteratively
\begin{itemize}
\item Use Eq.~\eqref{eq:meanfield_elips} to find the critical value of $\lambda$.
\item Find the groundstate of $H^\mathrm{GW,g}$ at this $\lambda,V$.
\item Update $\langle s^z_{j,j+1} \rangle,\langle s^+_{j,j+1} \rangle$.
\item Find the groundstate of $H^\mathrm{GW,m}$ at this $\lambda,V,\mu$.
\item Update $\langle\sigma^-_j\sigma^-_{j+1}\rangle, \langle\sigma^z_j\rangle$.
\end{itemize}
This procedure converges after a few iterations, yielding the transition line connecting smoothly to the point $\mu=\infty$, $\lambda=V/2$, as plotted in Fig.~\ref{fig:PD}.
When $\mu$ becomes sufficiently negative, $\gamma_x$ will tend to $4V$ and Eq.~\eqref{eq:meanfield_elips} will no longer have a real solution for $\lambda$ critical.

\bibliography{Reliability_iMPS}
\end{document}